\documentclass[10pt, a4paper, twocolumn]{article}

\usepackage{cite}
\usepackage{graphicx}
\usepackage{xcolor}
\usepackage[utf8]{inputenc}
\usepackage[english]{babel}
\usepackage[autostyle=true]{csquotes}
\usepackage{url}
\usepackage{footnote}
\usepackage{tabularx}
\usepackage{multirow}
\usepackage{booktabs}
\usepackage{listings}
\usepackage{enumitem}
\usepackage{amsfonts}
\usepackage{balance}
\usepackage{adjustbox}
\usepackage{hyperref}
\usepackage{amsmath}
\usepackage[all]{nowidow}
\usepackage[numbers,sort&compress]{natbib}
\usepackage[normalem]{ulem}
\usepackage[labelfont={bf}]{caption}
\usepackage{authblk}
\usepackage[centering,margin=1.5cm]{geometry}

\usepackage[capitalise,noabbrev]{cleveref}

\usepackage{subcaption}
\newcommand\free[0]{\vspace{1em}\noindent}
\newcommand{\proj}[1]{\textit{#1}}
\newcommand{\influx}[0]{\proj{InfluxDB}}
\newcommand{\victoria}[0]{\proj{VictoriaMetrics}}

\date{}
\begin{document}

\title{Using Microbenchmark Suites to Detect Application Performance Changes \thanks{This is the author copy of the paper published in IEEE Transactions on Cloud Computing \textsuperscript{\textcopyright}2022 IEEE (DOI: 10.1109/TCC.2022.3217947). Aside from the formatting, it is identical with the official IEEE version.}}

\author{Martin~Grambow$^1$,
				Denis~Kovalev$^1$,
				Christoph~Laaber$^2$,
				Philipp~Leitner$^3$,
        and~David~Bermbach$^1$}
				
\affil{$^1$Mobile Cloud Computing Research Group, TU Berlin \& Einstein Center Digital Future \\Berlin, Germany. E-mail: \{mg,dkv,db\}@mcc.tu-berlin.de}
\affil{$^2$Simula Research Laboratory \\Oslo, Norway. E-mail: laaber@simula.no}
\affil{$^3$Software Engineering Division, Chalmers $|$ University of Gothenburg \\ Gothenburg, Sweden. E-mail: philipp.leitner@chalmers.se}

\maketitle
				
\begin{abstract}
Software performance changes are costly and often hard to detect pre-release. 
Similar to software testing frameworks, either application benchmarks or microbenchmarks can be integrated into quality assurance pipelines to detect performance changes before releasing a new application version. 
Unfortunately, extensive benchmarking studies usually take several hours which is problematic when examining dozens of daily code changes in detail; hence, trade-offs have to be made. 
Optimized microbenchmark suites, which only include a small subset of the full suite, are a potential solution for this problem, given that they still reliably detect the majority of the application performance changes such as an increased request latency. 
It is, however, unclear whether microbenchmarks and application benchmarks detect the same performance problems and one can be a proxy for the other.

In this paper, we explore whether microbenchmark suites can detect the same application performance changes as an application benchmark. 
For this, we run extensive benchmark experiments with both the complete and the optimized microbenchmark suites of the two time-series database systems \influx{} and \victoria{} and compare their results to the results of corresponding application benchmarks. 
We do this for $70$ and $110$ commits, respectively. 
Our results show that it is possible to detect application performance changes using an optimized microbenchmark suite if frequent false-positive alarms can be tolerated.
\end{abstract}

\vspace{1em}


\section{Introduction}
\label{sec:introduction}

Performance issues in software systems should be identified and dealt with as early as possible.
Besides a poor user experience, performance issues can also occupy additional resources and result in major fixing efforts which all imply unpredictable additional costs~\cite{zaman_qualitative_2012, zaman_security_2011, chen_exploratory_2017}. 
Thus, performance changes should ideally be detected by a Continuous Integration and Deployment (CI/CD) pipeline immediately after a code change is checked in~\cite{bulej_repeated_2005, foo_mining_2010, grambow_continuous_2019, waller_including_2015, mostafa_tracking_2009, ingo_automated_2020, daly_creating_2021}.

For validating performance properties or adherence to Service Level Agreements (SLAs) such as a specific maximum latency or processing duration, software engineers often use benchmarking. 
Here, a system under test (SUT) is stressed with an artificial load, the requested values are measured, and these are then compared with the specification values, with the results of another alternative system, or with a previous version~\cite{bermbach_cloud_2017}. 
For detecting a performance change using benchmarking, there are two alternatives with different levels of granularity:
first, using application benchmarks, where the SUT is set up including all related components and stressed in an environment that mimics the production conditions (e.g., a database system running on a virtual instance is stressed by a client software which mimics the requests of thousands of users for half an hour)~\cite{bulej_repeated_2005, grambow_continuous_2019, binnig_how_2009, bermbach_benchfoundry_2017};
second, using microbenchmarks, which analyze individual functions at source code level and execute them repeatedly (e.g., a date conversion method is called a million times)~\cite{laaber_performance_2018, laaber_evaluation_2018}.
While the former method provides reliable results regarding application runtime implications, it is complex due to the setup and execution of the application benchmark.
Microbenchmarks, on the other hand, are simpler and less complicated to execute.
Nevertheless, microbenchmarks are unable to reliably detect all problems, as they do not take the integration of the respective functions or modules into the overall (production) system into account.
For a single (standalone) component, however, they might be used as a proxy for a complex application benchmark.

Nevertheless, neither benchmarking technique is currently suited to be executed on every code change due to the extensive execution durations of several hours as well as the resulting costs~\cite{laaber_evaluation_2018, silva_cloudbench_2013, grambow_continuous_2019, waller_including_2015, daly_industry_2019}.
Applying one of these two approaches to a large project with many application developers, hundreds of source code files, and multiple code changes per day would soon create a stack of benchmark tasks that would prevent fast-paced software development and integration of individual changes. 
Optimized microbenchmark suites containing only a small number of microbenchmarks can potentially solve this problem, because their execution is orders of magnitude faster, yet they also have to reliably detect application-relevant performance changes. 
A recent approach by~\citet{grambow_using_2021} proposes to optimize microbenchmark suites by removing redundancies within the suite and only executing practically relevant microbenchmarks, i.e., microbenchmarks which cover source code parts that are frequently used in production (represented by an application benchmark as baseline).

\free
\noindent
In this paper, we investigate to which extent application benchmarks and microbenchmarks detect the same performance changes, and if we can use a smaller, optimized microbenchmark suite as a proxy for application benchmarks.
To this end, we apply the optimization approach of~\citet{grambow_using_2021} in real world examples using the open source systems \influx{} and \victoria{} as case studies and execute application benchmarks, optimized microbenchmark suites, and complete microbenchmark suites against $70$ and $110$ commits, respectively. 

\free
\noindent
This paper makes the following main contributions: 

\begin{itemize}
\item A comprehensive benchmarking dataset of application benchmarks and microbenchmarks for series of successive code changes which is available openly.\footnote{\url{http://dx.doi.org/10.14279/depositonce-15532}}
\item A performance change detection approach for the resulting time series data which differentiates between performance jumps and trends as well as potential and definite performance changes.
\item An impact metric for quantifying the implications of individual microbenchmark results on application performance.
\item Empirical evidence showing that less expensive and faster optimized microbenchmark suites can be used as a proxy for application benchmarks in certain situations.
\end{itemize}

\free
\noindent
For the two evaluated systems, the results show that performance changes can be reliably detected by running an application benchmark for less than one hour and that a reduced and optimized microbenchmark suite can detect the same changes with less than ten microbenchmarks.
Our experiments identify nine true positive detections for optimized microbenchmark suites.
Nevertheless, our study also shows the limitations of the optimization approach and which type of performance issues cannot be identified with an optimized suite.
First, the optimized suite does not detect the application performance changes if the microbenchmarks do not cover the practically relevant code sections.
Second, performance changes related to the concrete runtime environment may not be detected.
Third, if a performance change is detected by a microbenchmark, its impact on application performance is hard to predict. 
The optimized suites hence often identify false positives.

\free
\noindent
Derived from our findings, we envision that a good continuous benchmarking strategy should, e.g., combine a fast and relevant optimized microbenchmark suite and a well-designed application benchmark.
While the optimized suite provides an early performance feedback for almost every code change, e.g., as part of a local build process or routine action which is triggered for each submitted code change, the regular runs of a well-designed application benchmark, e.g., once per day, ensures that the desired performance metrics are met. 
This allows developers to get quick performance feedback for each change, which allows them to adjust their changed code sections if necessary.
Because optimized suites may not or even cannot find all problems, a daily application benchmark run serves as backup to reliably detect the remaining ones and report them the next day.

\section{Background}
\label{sec:background}

This section introduces both benchmarking techniques, application benchmarks and microbenchmarks, and the approach for deriving optimized microbenchmark suites.

\free
\textbf{Application Benchmarks}

\noindent
Application benchmarks evaluate non-functional properties of an SUT by deploying the respective system and all related components in a production-like test or staging environment and stressing these with an artificial but realistic workload~\cite{bermbach_benchfoundry_2017,bermbach_cloud_2017}. 
Thus, application benchmarks are often seen as the gold standard, because they evaluate the respective systems using a realistic load in the actual runtime environment and, depending on the use case, also using specific load scenarios (e.g., increased visits and checkouts during the Christmas season).
A well-designed application benchmark can provide answers to many performance-related questions and also can be used to compare different versions of an SUT.
This is especially relevant for the context of this work, in which a dedicated benchmark step as part of a CI/CD pipeline is envisioned~\cite{grambow_continuous_2019, waller_including_2015}.
On the other hand, however, continuous benchmarking for early performance regression detection is expensive, complex, and time-consuming~\cite{bulej_repeated_2005, silva_cloudbench_2013}. 
Besides the setup and configuration of all relevant components, which can already take a considerable amount of time, all experiments have to run for a certain time and usually have to be executed several times to get reliable results, especially in cloud environments~\cite{laaber_performance_2018}.

\free
\textbf{Microbenchmarks}

\noindent
Microbenchmarks evaluate an SUT on function\footnote{We use the term \textit{function} to refer to any form of subroutine, no matter how they are called in the respective programming language.} level by executing individual functions multiple times.
These microbenchmarks typically call the respective function under test repeatedly with artificial parameter values for a specified duration and a specified number of iterations. 
Multiple microbenchmarks together form a microbenchmark suite, which is usually executed several times in a row to get reliable results. 

In contrast to application benchmarks, microbenchmarks are easier to set up and execute. 
Instead of (possibly) compiling, starting and configuring various components, it is usually enough to compile the corresponding code files and start the microbenchmark suite with the respective configuration.
Similar to unit tests, they can even be run inside the local development environment. 
On the other hand, microbenchmarks are usually considered less powerful because it is unclear whether they cover relevant parts of the production system or if detected performance changes will affect the production system~\cite{heger_automated_2013}, which is what we will investigate in this paper.

\begin{figure}[t]
	\centering
  \includegraphics[width=\linewidth]{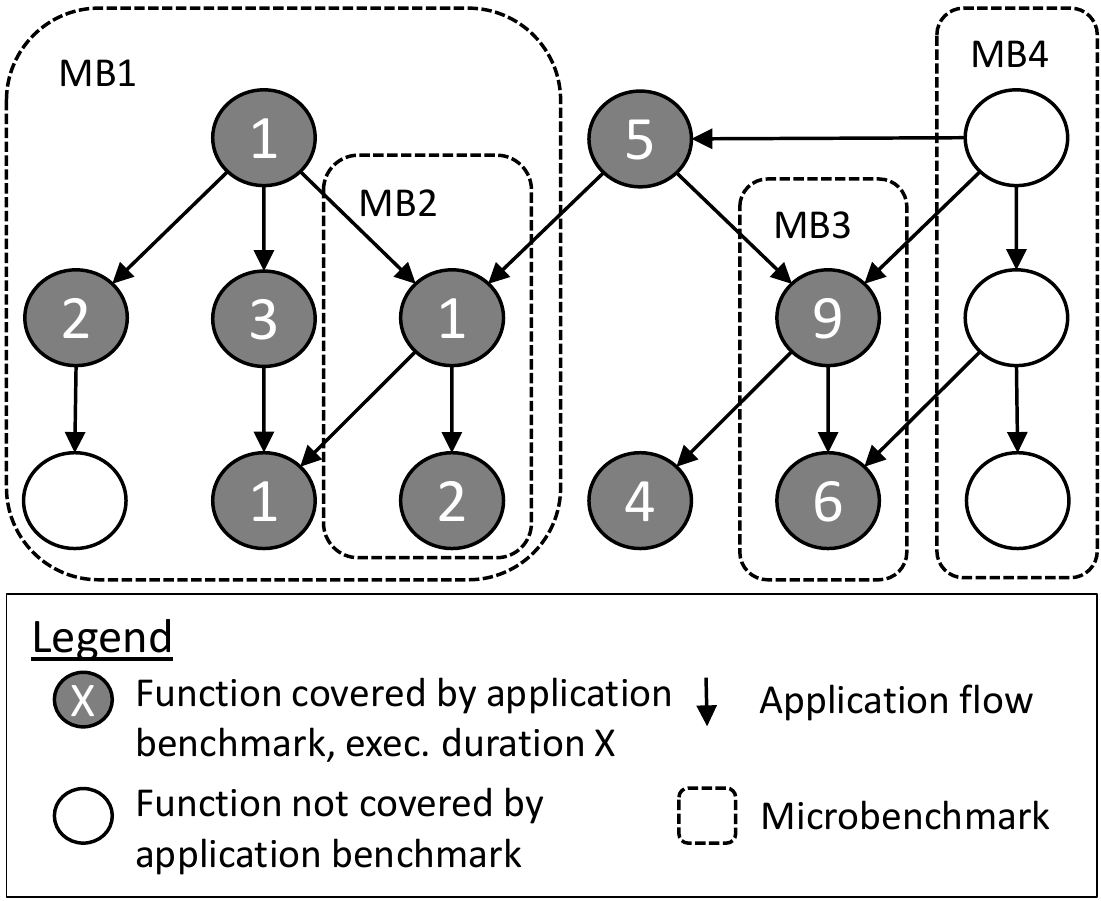}  
  \caption{\textbf{Strategy for optimizing microbenchmark suites:} A suite containing microbenchmarks (MB) 1 and 3 would cover $80\%$ of the application call graph. MB2 would not be included as all functions are already evaluated by MB1. MB4 does not evaluate any practically relevant functions.}
  \label{fig:approach}
	\vspace{-1em}
\end{figure}

\free
\textbf{Removing Redundancies in Microbenchmark Suites}

\noindent
Especially in large software projects with hundreds or thousands of microbenchmarks, a complete suite execution can take several hours, making the evaluation for every code change impractical. 
Thus, there are various approaches to optimize microbenchmark suites~\cite{de_oliveira_perphecy_2017} and also to detect performance problems using call graphs~\cite{mostafa_tracking_2009, javed_perfci_2020}. 
Application call graphs represent the individual methods and functions of an application as nodes and their respective calls to each other as edges. 
For this study, we use an approach that removes redundancies in a microbenchmark suite based on an application benchmark call graph and includes only practically relevant microbenchmarks, i.e., microbenchmarks evaluating functions that are actually used in production~\cite{grambow_using_2021} (see \cref{fig:approach}).
Here, a pre-recorded call graph from an application benchmark, which mimics the production behavior, is used as a baseline and compared with the call graphs of the respective microbenchmarks. 
The approach leverages a greedy heuristic and iteratively appends the microbenchmarks with the respective largest common overlap within the graphs to the optimized microbenchmark suite until no more new nodes are introduced.
This excludes microbenchmarks which evaluate the same code sections and microbenchmarks which do not evaluate practically relevant code sections from the suite, thus reducing the number of benchmarks and significantly shortening the overall suite execution duration.
The final optimized suite then solely consists of microbenchmarks that actually evaluate functions that are relevant in a production environment (which is represented by the application benchmark call graph).

\section{Study Design}
\label{sec:study}

Applying and analyzing both benchmark types in realistic setups requires large software projects as study objects and a long commit history.
Moreover, it must be possible to run a standardized application benchmark and there must be an extensive microbenchmark suite.
Finally, to apply the removal of redundancies within the suite, it must be possible to trace the call graphs during the respective benchmark.

In the following sections and experiments, we therefore use two open source time-series database systems (TSDB) as study objects, namely \victoria{} and \influx{}. 
Both database systems are written in Go (which allows us to trace the call graphs without major modifications), come with extensive histories of code changes, and have comprehensive microbenchmark suites.
Moreover, there are application benchmarks for both TSDBs.
Because it is infeasible to examine the entire development cycle over a period of several years, we examine a smaller sample of successive code changes, spanning several months, to simulate a realistic long-term use of all benchmarking techniques.

\free
To study to which degree performance changes can be detected with an optimized microbenchmark suite, we initially run application benchmarks for both TSDBs to detect all application performance changes for the production environment. 
Next, we optimize the respective microbenchmark suites using recorded application benchmark call graphs as reference and run the optimized suite for every code change as well to check whether the optimized suites are capable of detecting the same performance changes.
Finally, we also run the complete microbenchmark suites to quantify the degradation of detection quality caused by relying only on the optimized microbenchmark suite.

\free
An optimized suite capable of detecting relevant performance changes can be embedded in a cloud-based CI/CD pipeline.
To mimic this realistic setup as closely as possible, we therefore use cloud-based virtual machines (VMs) that are created and configured for every experiment run from scratch. 
To minimize performance variation between different instances and due to random effects such as noisy neighbors in the cloud environment, we adapt and apply recent best practices in each benchmarking discipline to acquire reliable measurement results:
we use the Duet Benchmarking technique~\cite{bulej_initial_2019, bulej_duet_2020} for application benchmarks and Randomized Multiple Interleaved Trials (RMIT)~\cite{abedi_conducting_2015, abedi_conducting_2017} for execution of microbenchmarks. 
For all experiments, we use a hardware setup that is similar to the cloud experiment setups in related studies~\cite{laaber_evaluation_2018, grambow_continuous_2019, grambow_is_2019, grambow_benchmarking_2020, borhani_wpress_2014}.
We run all experiments on e2-standard-2 Google Cloud instances in the europe-west3 region with 2 virtual CPUs and 8 GB RAM, local SSD storage, running Ubuntu 20.04 LTS.

\subsection{Study Objects}

TSDBs are optimized for storing sequences of time-stamped data, analyzing these sequences for specified time frames, and thus detecting trends and anomalies.
Usually, values arrive in order and are appended to an existing time series whereas delayed values are inserted less frequently, e.g., due to network fluctuations.
Furthermore, existing values are updated rarely and many TSDBs support grouping queries based on tagging~\cite{dunning_time_2014}. 

\begin{table}[t]
\centering
\caption{\textbf{Study objects and meta information.} Both TSDBs can be evaluated using application and microbenchmarks.}
\begin{adjustbox}{max width=\columnwidth}
\begin{tabular}{@{}llrr@{}}\toprule
	Project &  & \victoria{} & \influx{} \\
	\midrule
	Go files &  & 2,088 & 1,653 \\
	Lines of Go Code &  & 742,191 & 520,716 \\
	\midrule
	Branch / Release &  & master & influx2.0\\
	Start of Evaluation Period &  & Mar 1, 2021 & Jan 1, 2021\\
	End of Evaluation Period &  & May 31, 2021 & May 14, 2021 \\
	Number of Commits &  & 70 & 110 \\
	Number of Microbenchmarks &  & 177 & 426 (109)\\
\bottomrule
\end{tabular}
\end{adjustbox}
\label{tbl:studyObjects}
\vspace{-1em}
\end{table}

\free

Since the first version of \victoria{} was released in 2018, more than $80$ contributors have created more than $2,000$ files and made more than $2,800$ commits as of May 31, 2021. 
In our experiments, we study the most recent $70$ commits at the time of running the experiments, i.e., between March 1 and May 31, 2021, which merge a pull request into the master branch, in more detail.
In this period, the complete microbenchmark suite consists of $177$ executable microbenchmarks in total, including all parameterized factors.

\influx{} squashes individual fixes and features in single commits. 
\influx{} version 2.0, which we investigate further, accumulated over $34,000$ commits in more than $1,600$ files from $422$ contributors up to May 14, 2021. 
For our detailed study, however, we examine the most recent $110$ commits at the experimentation phase in the time frame between Jan 1 and May 14, 2021.
In contrast to \victoria{}, the complete microbenchmark suite of \influx{} does not remain constant for our evaluation period, but decreases from $426$ microbenchmarks in the beginning to $109$ at the end (we address and discuss this in \cref{sec:results,sec:discussion}).
\Cref{tbl:studyObjects} gives an overview of both study objects and the respective commits that we studied.

\subsection{Application Benchmarks}
\begin{figure}[t]
	\centering
  \includegraphics[width=0.7\linewidth]{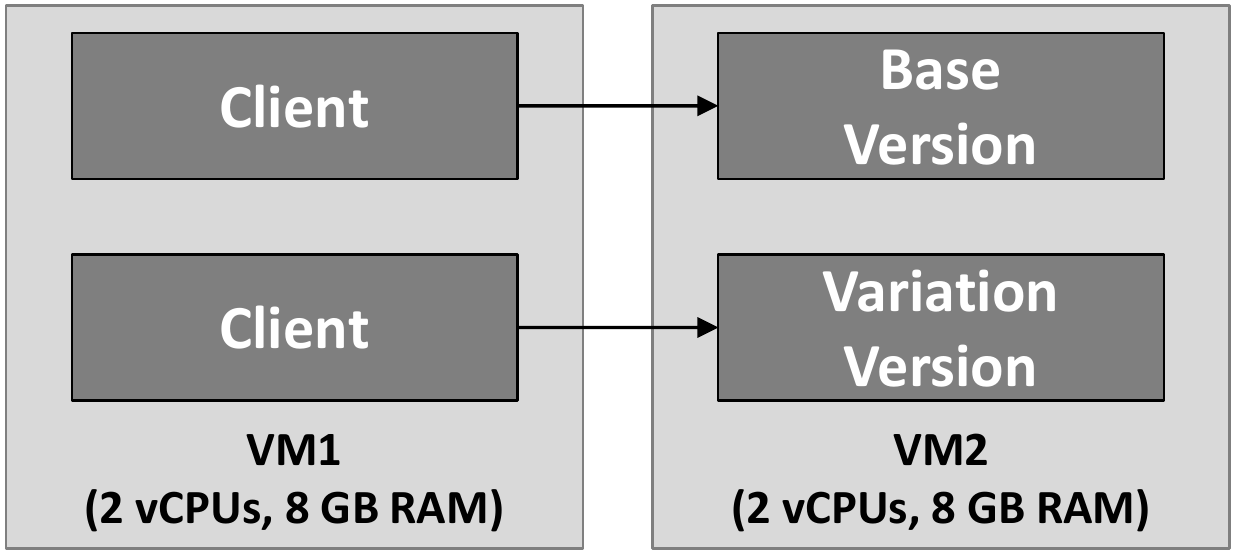}  
  \caption{\textbf{Application benchmark setup.} To compare the performance of the first version (base) with the current version (variation) of the SUT in the respective evaluation period, we deploy both variants on the same VM and benchmark both simultaneously.}
  \label{fig:appSetup}
	\vspace{-1em}
\end{figure}

For the application benchmarks, we use two cloud VMs in each experiment, one for the client sending the load and one for the respective SUT. 
Moreover, we adapt the Duet Benchmarking technique~\cite{bulej_initial_2019, bulej_duet_2020} and set up two versions of the respective TSDB as Docker containers on the same VM: the base version and the variation (see \cref{fig:appSetup}).
We then use a pre-generated workload and start two benchmark clients simultaneously, one targeting the base version's port and one targeting the variation's port.
Thus, the performance of the two variants can already be compared at three experiment repetitions, because both SUT versions are exposed to the same random factors at the same time.

\free
The application benchmark workload is based on the \mbox{\textit{DevOps}} use case in the Time Series Benchmark Suite (\textit{TSBS}\footnote{\url{https://github.com/timescale/tsbs}}), which in turn is based on the client \textit{influxdb-comparisons}.\footnote{\url{https://github.com/influxdata/influxdb-comparisons}}
We use the \textit{TSBS} client to benchmark \victoria{} and \textit{influxdb-comparisons} to benchmark \influx{}, but have extended both clients to report latency values of inserts and queries separately.\footnote{\url{https://github.com/martingrambow/benchmarkStrategy}}
The \textit{DevOps} use case simulates a server farm in which a specified number of servers sends utilization data (e.g., CPU and RAM) to the TSDB in a specified interval. 
After a first phase in which the data is inserted into the database, a second phase continues with simple queries, and a final phase with more complex group-by queries completes the experiment.
We adjusted the number of simulated servers, the sending interval, the total simulated duration, and the number of respective queries to the specific SUT in a way that the client instance is below $50\%$ utilization and the SUT instance is almost always fully utilized (see \cref{tbl:workload} for all workload details). 
Furthermore, we repeat each experiment at least three times using fresh VM instances to ensure reproducibility. 

\begin{table}[t]
\centering
\caption{\textbf{Workload parameters.} The workload for each TSDB differs to ensure full utilization of the respective SUT.}
\begin{adjustbox}{max width=\columnwidth}
\begin{tabular}{@{}llrr@{}}\toprule
	Project &  & \victoria{} & \influx{} \\
	\midrule
	Number of Simulated Servers &  & $800$ & $100$\\
	Sending Interval &  & $60s$ & $60s$\\
	Simulated Duration &  & $72h$ & $168h$\\
	Number of Insert Clients &  & $4$ & $10$\\
	Batch Size &  & $400$ & $60$\\
	Number of Batches & & $259,200$ & $113,400$ \\
	Number of Simple Queries &  & $8,640$ & $1,008$ \\
	Number of Group-By Queries &  & $1,440$ & $168$\\
	Number of Query Clients &  & $10$ & $10$\\
\bottomrule
\end{tabular}
\end{adjustbox}
\label{tbl:workload}
\vspace{-1em}
\end{table}

\free
For our interpretation of results, we have to consider two aspects: 
first, as in almost all application benchmark experiments, the first few measurements must be considered as part of a warm-up phase and should be discarded~\cite{bermbach_cloud_2017};
second, due to the duet benchmarking, the last measurements should be removed as well. 
If one of the two evaluated versions has better performance, then the respective benchmark run will also finish earlier than the other one, which will then release resources on the experiment VM. 
The other container running the slower version then has access to additional resources and speeds up, which leads to wrong measurements.
After some initial experiments comparing the first and last commit state of our study periods to determine the expected overall performance change, we chose to disregard the first $5\%$ and the last $20\%$ in the measurement series of each application benchmark run. 

\subsection{Microbenchmarks}

To remove redundancies in the microbenchmark suite, we initially execute and trace an application benchmark against the first commit of the evaluation period and create an application call graph, which serves as the reference for the optimization algorithm. 
Next, we execute and trace the full microbenchmark suite for the same commit and generate the call graph for each microbenchmark.
With both inputs, the reference application graph and the microbenchmark graphs, we then determine the practical relevance of the microbenchmark suites, remove redundancies, decide which microbenchmarks to include in the optimized suite, and run the optimized microbenchmark suites for every commit in the respective evaluation period~\cite{grambow_using_2021}.
Finally, to rate the improvements and back-test the optimization, we also run the full microbenchmark suites for every 5th commit in the evaluation period.
Running all microbenchmarks for every commit in practice is unrealistic due to the high costs of execution.
An execution for every 5th commit is a trade-off between a very detailed analysis and a long execution time as well as the corresponding monetary costs.
We believe that this is fine-grained enough to detect relevant changes and, in case there are anomalies, to further evaluate the relevant benchmark for the intermediate changes.

\free
To mimic the usage of the microbenchmark suites in CI/CD pipelines, which compare a new version with an older commit state, we benchmark both versions on the same VM using RMIT time-shared execution~\cite{abedi_conducting_2015, abedi_conducting_2017}. 
Here, to counteract infrastructure variation, we randomize the execution order of each suite and run each microbenchmark for both versions successively.
To reduce the influence of the microbenchmarks on the performance of the following ones and to make sure that these effects are not systematic, we also randomly vary which microbenchmark version (base or variation) is executed first.
Adapted from the configurations used by~\citet{laaber_predicting_2021} and~\citet{chen_exploratory_2017}, we repeat each of our microbenchmark experiments three times on fresh VMs (instance run), run each suite three times (suite run), and call each benchmark five times (iteration) for one second each (duration).
In total, thus, there are $45$ measurements per microbenchmark per commit, each comprising many benchmark function calls.

\subsection{Analysis}
\label{sub:analysis}

We analyze the results of the respective benchmarks as follows:

\free
\textbf{Compared Versions}

\noindent
For all our experiments, we fix the base version to the first commit in the evaluation period and iterate over the commits as variation version.
Thus, we always compare the current variation with the initial one. 
Nevertheless, because the results are transitive, the performance changes can be visualized as a pseudo-continuous graph.

\free
\textbf{Median Performance Change}

\noindent
To actually compare the versions, i.e., to decide which version performs better, we use the median value of all measurements. 
For the application benchmark, we use the median latency of all measured latency values for the respective query type.
For microbenchmarks, we use the median execution duration of each microbenchmark of the $45$ measurements ($3$ instance runs $*$ $3$ suite runs $*$ $5$ iterations).
Finally, we calculate the relative change by comparing the median value of the base version with the median of the variation (e.g., if the median latency increases from $100ms$ to $110ms$, a query takes $10\%$ longer).

\begin{figure}[t]
\centering
\begin{subfigure}{.48\linewidth}
  \centering
  \includegraphics[width=\linewidth]{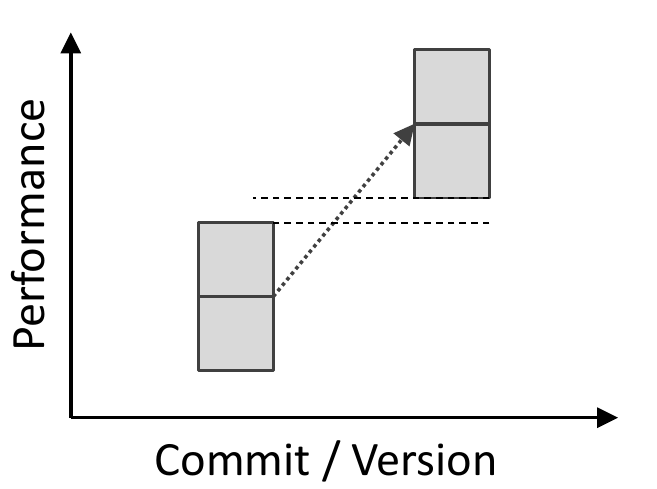}  
  \caption{Definite Change}
  \label{fig:definite}
\end{subfigure}
\begin{subfigure}{.48\linewidth}
  \centering
  \includegraphics[width=\linewidth]{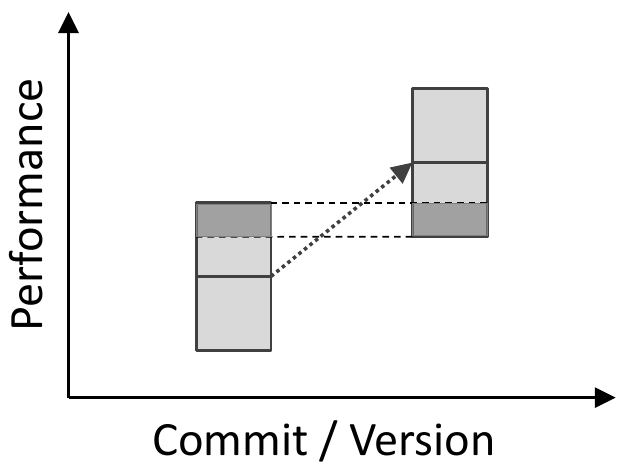}  
  \caption{Potential Change}
  \label{fig:potential}
\end{subfigure}
\caption{\textbf{Intensity classification.} Experiments in cloud environments can show a large variance. We therefore classify detected
changes based on the $99\%$ CI as definite or potential.}
\label{fig:perfChanges}
\vspace{-1em}
\end{figure}

\free
\textbf{Confidence Intervals}

\noindent
Moreover, to determine the confidence interval (CI) of a performance change, we use a bootstrapping methodology that implements hierarchical random re-sampling with replacement~\cite{kalibera_quantifying_2020}.
For the microbenchmarks, we draw $10,000$ random samples of $45$ values each from the measurements\footnote{Due to the replacement, values can be drawn multiple times and this precisely maps the actual distribution of the measurements.}, determine the median value in each sample, and use the top and bottom $\alpha=0.5\%$ of the resulting ordered set of the medians as the $99\%$ CI. 
For the application benchmarks, we adapt the sample size to the number of requests for the respective request type, draw $10,000$ samples, and determine the CI in the same way.

\free
\textbf{Definite and Potential Performance Changes}

\noindent
A wide CI of an experiment implies that the concrete performance change of a (micro-)benchmark cannot be clearly quantified and that the individual benchmark is unstable. 
Thus, we refer to the width of a confidence interval as \textbf{instability}.
The smaller this instability is, the better and more precisely it is possible to detect performance changes.
On the other hand, a wide CI implies that it is only possible to detect large performance changes, because overlapping CIs of the respective experiments do not allow us to draw precise conclusions.
Thus, we classify the detected performance changes as ($99\%$) \textbf{definite} (no overlap) and \textbf{potential} (overlapping CIs) performance changes (see \cref{fig:perfChanges}). 

\begin{figure}[t]
	\centering
  \includegraphics[width=\linewidth]{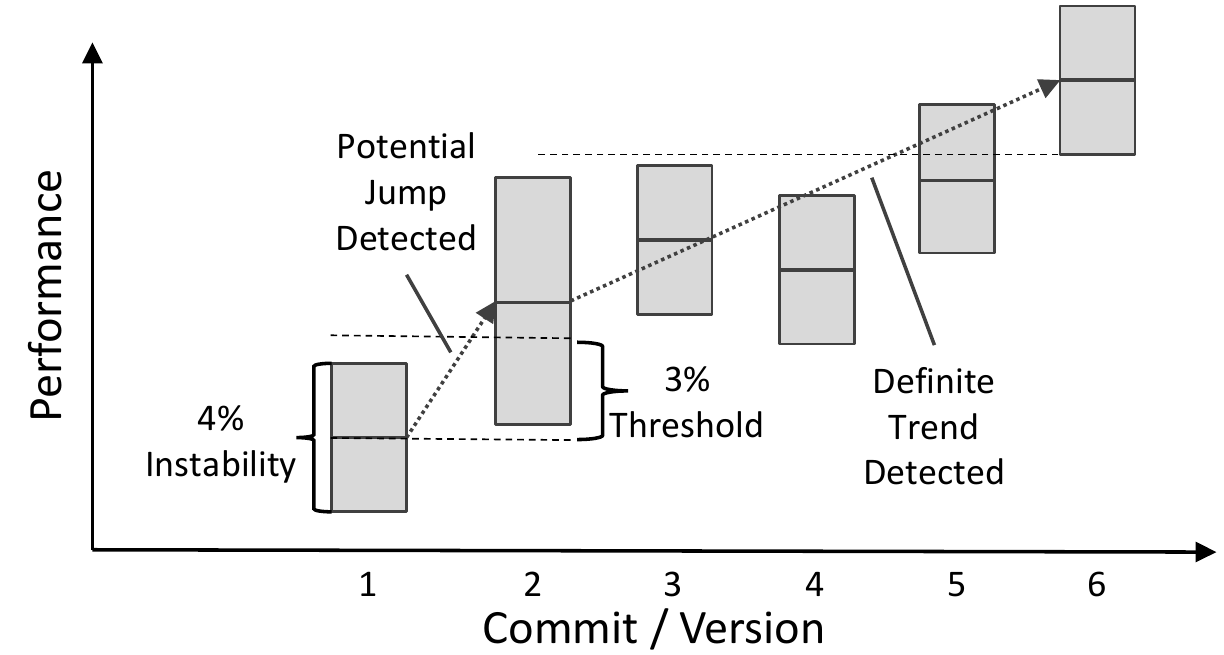}  
  \caption{\textbf{Type classification.} While the jump detection identifies performance changes in two successive code changes, the trend detection considers a series of commits. The detection threshold adapts dynamically to the previous instability measurements. Thus, the detection threshold for the 3rd commit would increase because of the larger instability in the second one.}
  \label{fig:detection}
	\vspace{-1em}
\end{figure}

\free
\textbf{Jump and Trend Detection}

\noindent
We adapt two basic threshold-based algorithms by~\citet{grambow_continuous_2019} to decide if we found a relevant performance change.
For this study, we use (i) a \textbf{jump} detection algorithm to identify individual commits that introduce performance changes, and (ii) a \textbf{trend} detection algorithm to detect performance trends in a series of ten commits. 
In our study, however, we extend the static threshold and use a dynamic one that constantly adjusts to $75\%$ of the instability of previous measurements (see \cref{fig:detection}).
Using $75\%$ of the CI width is a trade-off between many false positives ($50\%$) and potentially many false negatives ($100\%$).
Taking half the CI width could create false positive alarms in the change point detection, as the median performance change might just randomly fluctuate into the respective CI.
For example, in \cref{fig:detection}, using $50\%$ of the instability in commit 1 corresponds to a $2\%$ threshold, which leads to the median change in commit 2 to be exactly on the CI boundary of commit 1. 
Using the full CI width ($100\%$) would only detect changes larger than the referenced instability, e.g., $4\%$ for commit 2 in \cref{fig:detection}. 
For both of our projects studied, using a $75\%$ dynamic threshold provides a good balance between false-alarms and (potentially) undetected performance changes.
Nevertheless, this threshold parameter is project-specific, especially if the median performance value is not centered in the respective CIs but is shifted to either side.

Code changes that stabilize the measurements will thus narrow the CI automatically (or the other way around) and random cloud fluctuations during the complete experiment series will automatically be considered in the analysis.
Moreover, because we do not consider small performance changes as relevant in our evaluated projects, we also set a minimum threshold of $1\%$, similar to what best practice suggests~\citep{georges_statistically_2007}.
In other projects, however, even smaller changes may also be relevant and this value would have to be adjusted. 
Finally, our dynamic detection algorithms require an initial threshold that is close to the expected value. 
If the difference is too high at first, either many false alarms would be triggered (small initial threshold) or relevant changes would not be detected (large initial threshold). 
Nevertheless, after the threshold has been continuously adjusted across several code changes (in our case ten), the detection mechanisms are adjusted to the respective instability.

\free
\textbf{Reference Impact}

\noindent
Running optimized microbenchmark suites using bootstrapping analysis and dynamic thresholds will detect multiple potential and definite performance changes for several microbenchmarks.  
These results, however, cannot be directly linked to a request type in the application benchmark or allow other direct conclusions.
For example, if there is a definite performance drop of $5\%$ in a microbenchmark, this drop cannot be directly linked to application performance (e.g., slower queries).
To link a respective microbenchmark that detected a performance change to the application benchmark, we therefore use a \textbf{reference impact} value which is the sum of the execution durations of the overlapping functions in the reference application benchmark (see \cref{fig:approach}).
The key idea behind this is that a microbenchmark whose covered functions in the application benchmark have a smaller total execution duration (e.g., $10$ for MB1) will have less impact on overall application performance than another microbenchmark covering functions with a larger total application benchmark execution duration (e.g., $15$ for MB3).
We call this metric reference impact because it refers to the recorded application benchmark call graph and not to the performed microbenchmark experiment.
\section{Results}
\label{sec:results}

We first report the results of the application benchmarks and use them as \textquote{ground truth} for the optimization algorithm (\cref{subsec:results-app}). 
Next, we investigate whether the optimized microbenchmark suite can detect the same performance changes with less effort (\cref{subsec:results-opt}). 
To quantify the improvements and to verify that the complete microbenchmark suite is not a better proxy for detecting application performance changes, we also execute the complete suite for every 5th commit in our evaluation period (\cref{subsec:results-micro}). 
Finally, we derive implications combining all information (\cref{subsec:results-impl}).

\begin{table}[t]
\centering
\caption{\textbf{Result instability in A/A benchmarks.} All CIs are close to the $0\%$ value but insert requests to \victoria{} and queries to \influx{} show a larger instability.}
\begin{adjustbox}{max width=\columnwidth}
\begin{tabular}{@{}lrr@{}}\toprule
	& \multicolumn{2}{c}{Instability (99\% CI)} \\
	Project & \multicolumn{1}{c}{\victoria{}} & \multicolumn{1}{c}{\influx{}} \\
	\midrule
	Inserts & $6.27\%$ $[-2.95 ; 3.32]$ & $0.88\%$ $[-0.71 ; 0.17]$\\
	Simple Queries & $1.66\%$ $[-1.25 ; 0.41]$ & $3.37\%$ $[-1.36 ; 2.01]$\\
	Group-By Queries & $1.71\%$ $[-0.79 ; 0.92]$ & $2.11\%$ $[-0.82 ; 1.29]$\\
\bottomrule
\end{tabular}
\end{adjustbox}
\label{tbl:aatests}
\vspace{-1em}
\end{table}

\begin{figure*}[t]
\centering
\begin{subfigure}{.32\textwidth}
  \centering
  \includegraphics[width=\linewidth]{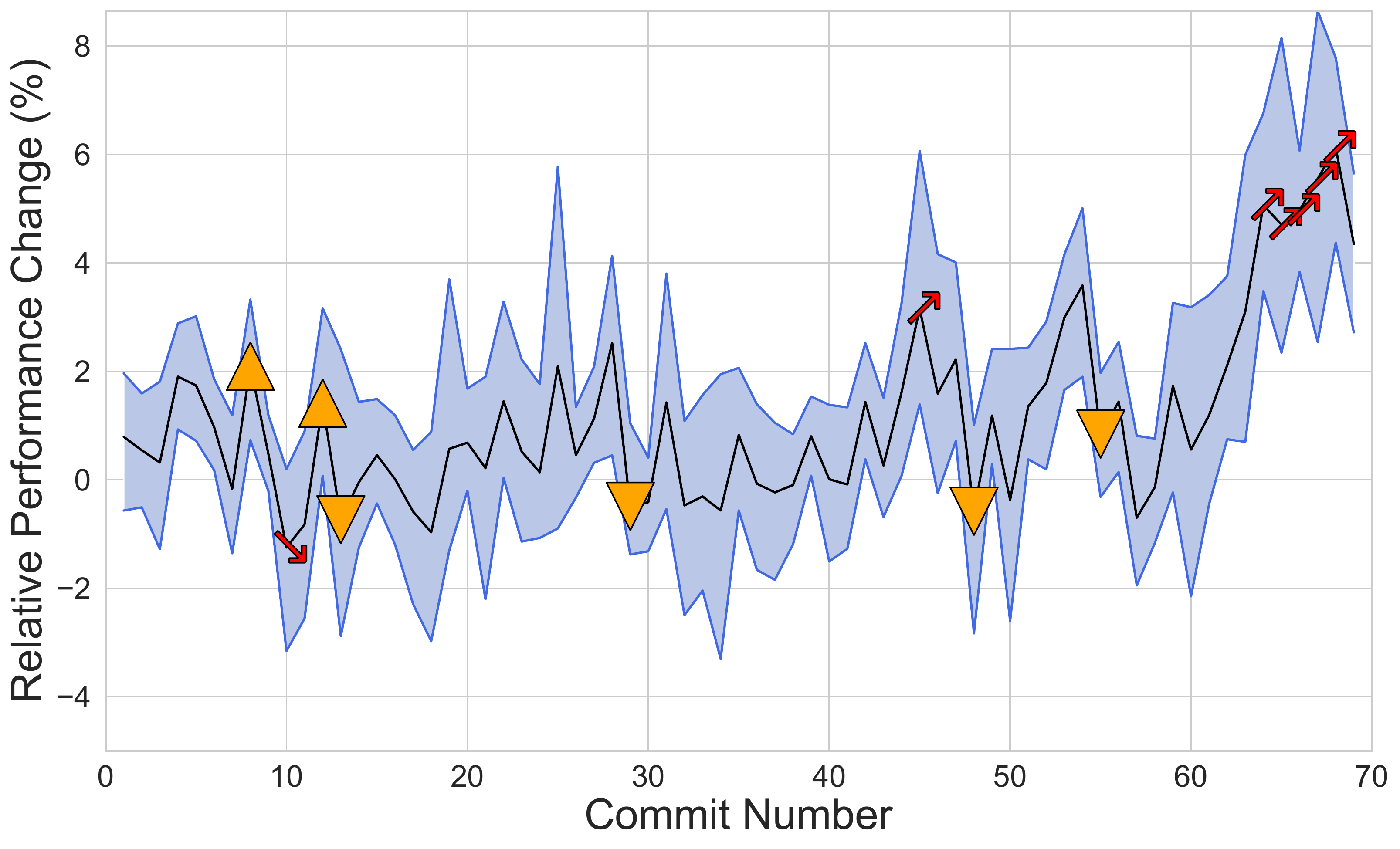}  
  \caption{Inserts}
  \label{fig:vm_inserts}
\end{subfigure}
\begin{subfigure}{.32\textwidth}
  \centering
  \includegraphics[width=\linewidth]{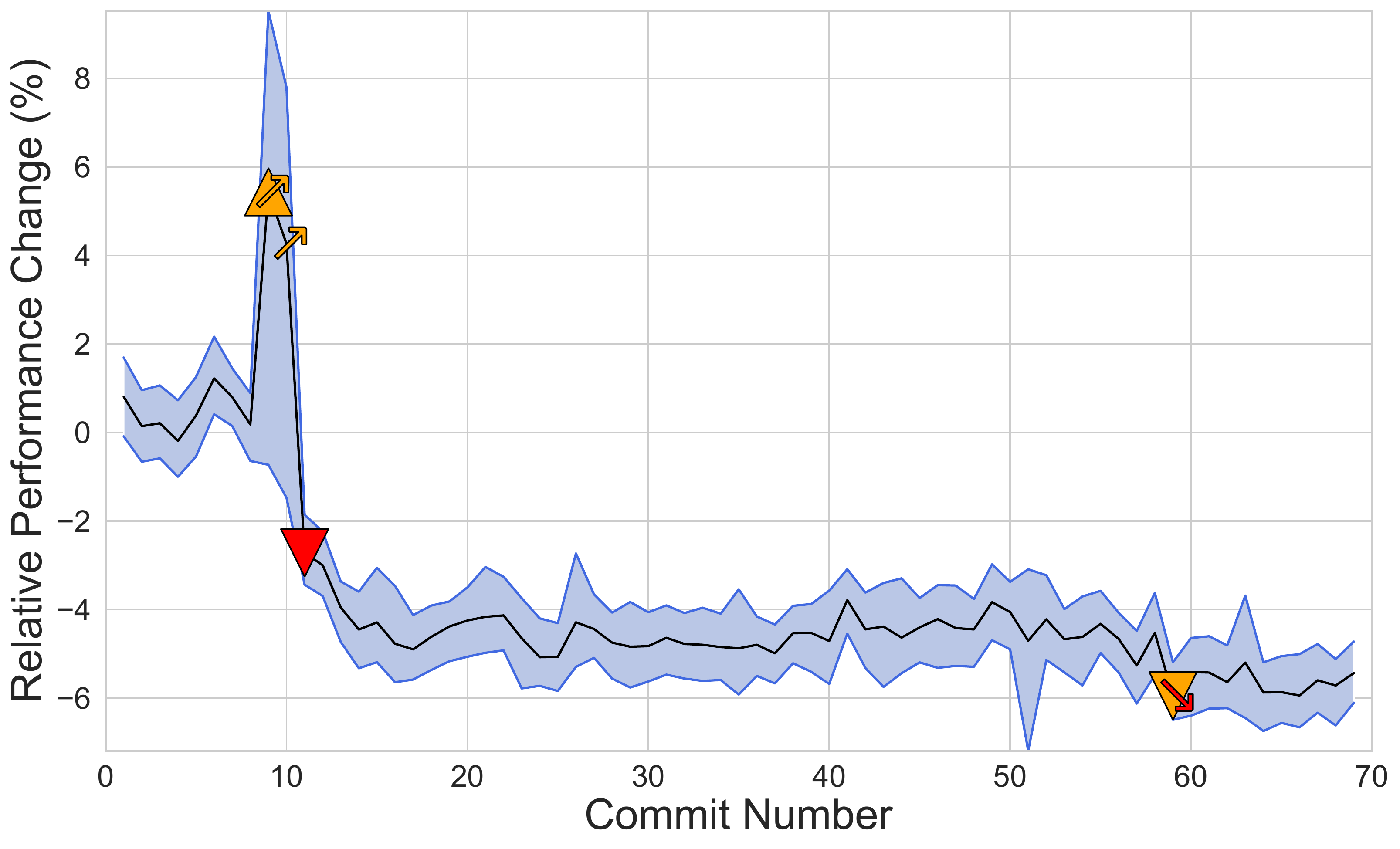}  
  \caption{Simple Queries}
  \label{fig:vm_query}
\end{subfigure}
\begin{subfigure}{.32\textwidth}
  \centering
  \includegraphics[width=\linewidth]{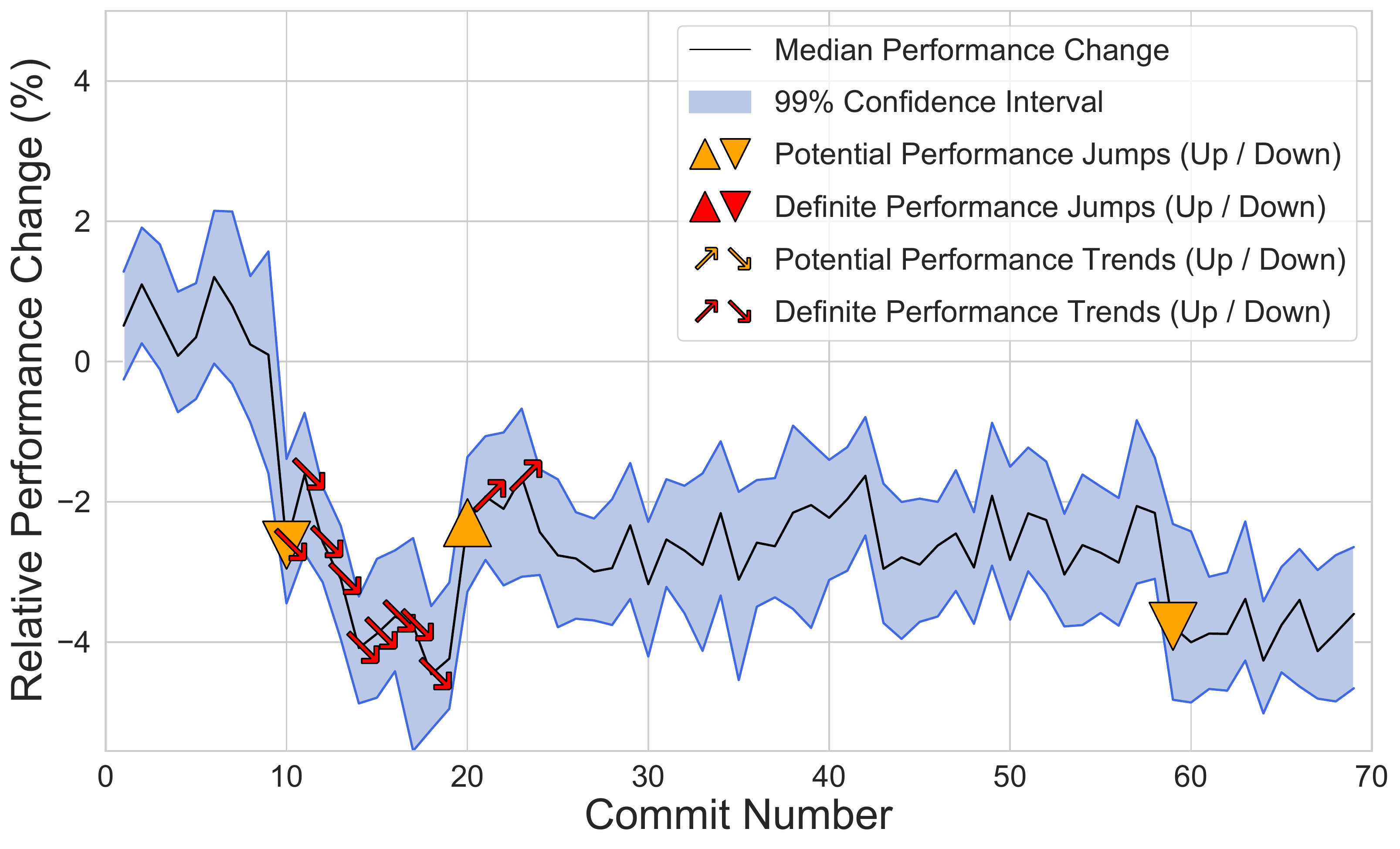}  
  \caption{Group-By Queries}
  \label{fig:vm_group}
\end{subfigure}
\caption{\textbf{Application benchmark results for \victoria{}, negative values show an improvement.} There is (i) a definite negative performance trend in the last commits of our evaluation period for inserts, (ii) a definite positive trend for both query types from commit $10$ to $20$ which is followed by (iii) a negative trend for group-by queries. Finally, there is (iv) a positive trend for both query types around commit $60$.}
\label{fig:app_vm}
\vspace{-1em}
\end{figure*}

\begin{figure*}[t]
\centering
\begin{subfigure}{.32\textwidth}
  \centering
  \includegraphics[width=\linewidth]{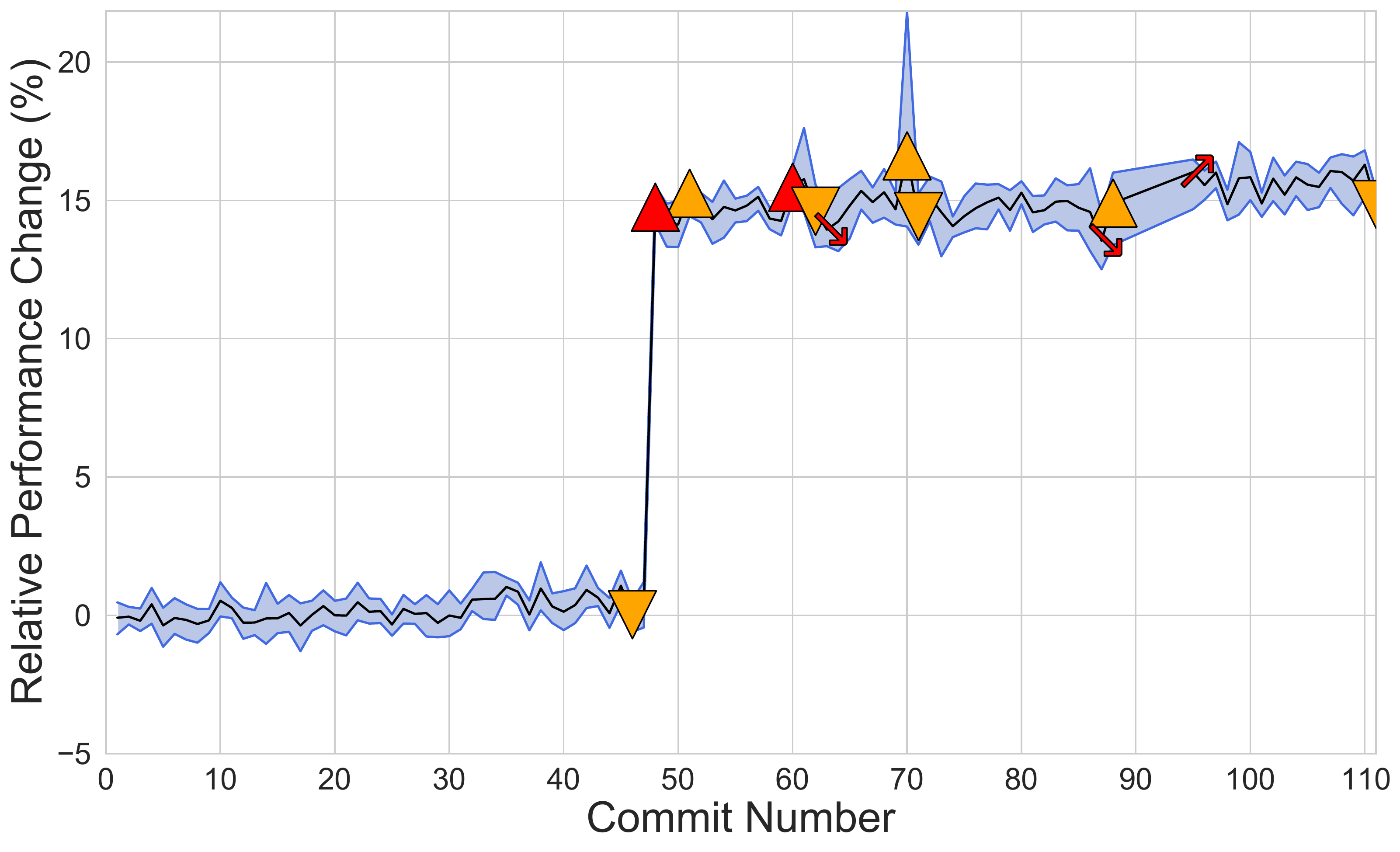}  
  \caption{Inserts}
  \label{fig:influx_inserts}
\end{subfigure}
\begin{subfigure}{.32\textwidth}
  \centering
  \includegraphics[width=\linewidth]{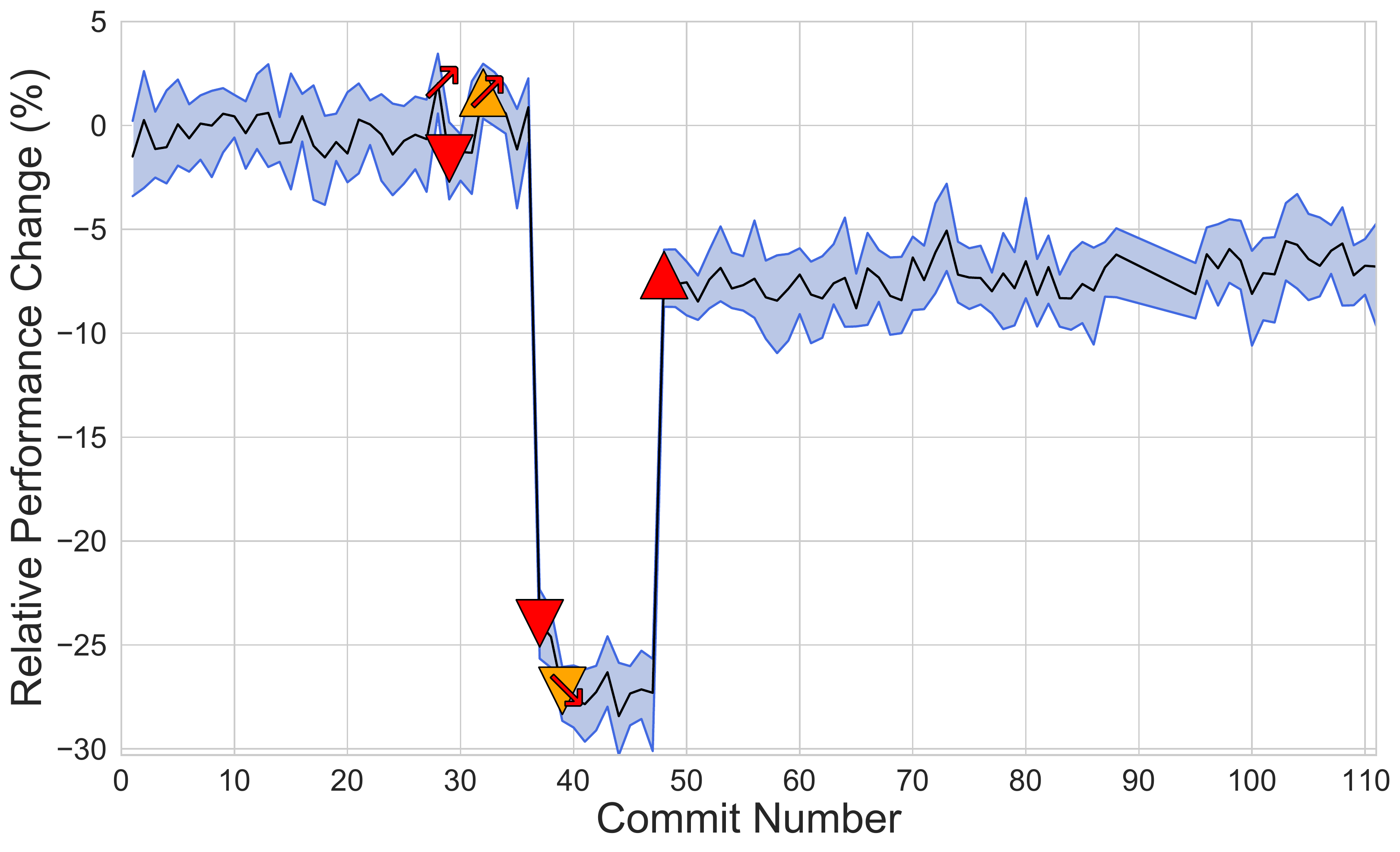}  
  \caption{Simple Queries}
  \label{fig:influx_query}
\end{subfigure}
\begin{subfigure}{.32\textwidth}
  \centering
  \includegraphics[width=\linewidth]{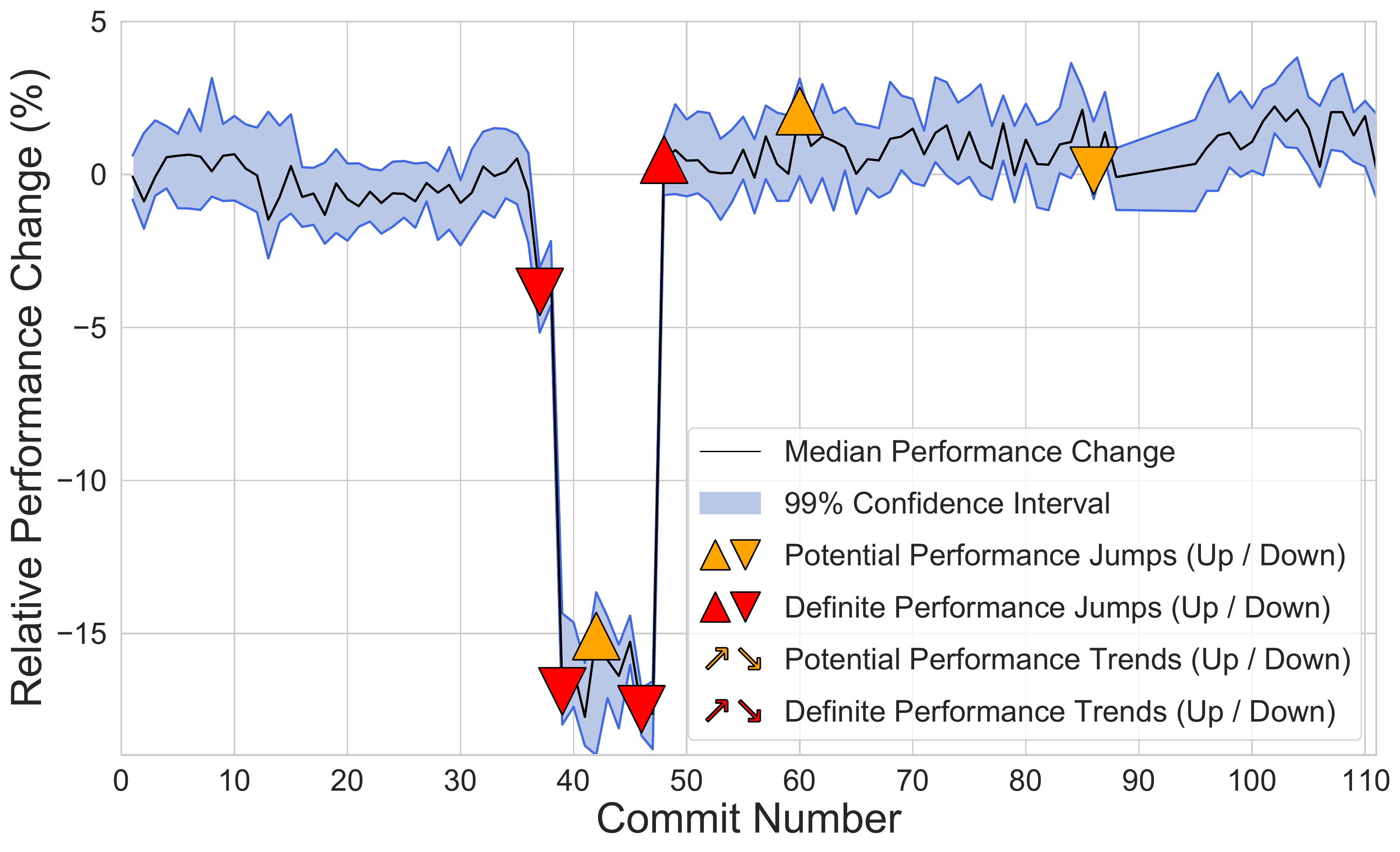}  
  \caption{Group-By Queries}
  \label{fig:influx_group}
\end{subfigure}
\caption{\textbf{Application benchmark results for \influx{}, negative values show an improvement.} There are two large definite performance jumps at (i) commit $48$ for all request types and at (ii) commit $37$ for both query types. Moreover, there are several (potential) jumps and trends for all request types.}
\label{fig:app_influx}
\vspace{-1em}
\end{figure*}

\subsection{Application Benchmarks\label{subsec:results-app}}

First, to verify that our results are reliable and correct, we use five repeated A/A benchmarks which compare the first commit as the base version with itself as the variation version.
Ideally there should not be any performance change, the CIs should be narrow, and around the $0\%$ value. 
\Cref{tbl:aatests} reports the respective CIs derived from the bootstrapping method for each query type and SUT. 
All CIs straddle $0\%$, which implies no detected performance change. 
Nevertheless, especially the wide CI for inserts in \victoria{} also implies that we can not reliably detect definite performance changes smaller than $6\%$.
Based on the respective CI (reported in \cref{tbl:aatests}), we set the initial detection thresholds for both change point detection algorithms to $\approx 75\%$ of the instability value or $1\%$ (see \cref{sub:analysis}, Jump and Trend Detection): 
except for the inserts in the case of \victoria{} ($5\%$) and the two query types in the case of \influx{} ($3\%$ and $2\%$), we thus set all the initial detection threshold values to $1\%$.

\free
\Crefrange{fig:vm_inserts}{fig:vm_group} show the relative performance history and the detected performance changes for insertions, simple queries, and group-by queries against \victoria{}.
A positive percentage value indicates that the respective request latency has increased. 

Due to the non-deterministic setup of the internal data structure in \victoria{}, there is a large instability for inserts.
To overcome this obstacle, we split the initial insertion phase in half, copied the data from the base version container after the first half of insertions, and replaced the data in the variation container with this copy.
The second part of the inserts is thus based on the same data structure and the non-determinism has a smaller effect on the result.
To ensure that the queries are also based on the same underlying data structure, we repeat this step after the second half of inserts.
Despite this instability, \Cref{fig:vm_inserts} clearly shows that the insertions become significantly slower in the overall sequence of $70$ commits and the change detection algorithm also detects this definite trend in the last ten commits.
While simple queries improve during our study period by almost $6\%$, the performance history of group-by queries shows more change points. 
Starting with commit $10$, the performance of complex group-by queries improves initially, then degrades from commit $20$ to $23$, and improves again with commit $59$.

\free
\Crefrange{fig:influx_inserts}{fig:influx_group} show the detected performance jumps and trends for \influx{} along with the measured relative performance history.
Besides several detected (potential) jumps and trends, both query types are significantly improved through commit $37$ and there is one major drop for all request types introduced with commit $48$.

The corresponding commit message for commit $37$, \textquote{feat(query/stdlib)~\cite{influxdb}: promote schema and fill optimizations from feature flags}, signals that a new feature successfully speeds up simple queries by around $25\%$ and group-by queries by around $15\%$. 
This improvement, however, is reversed in commit $48$ through the activation of profiling. 
The corresponding commit message for commit $48$, \textquote{feat(http): allow for disabling pprof}~\cite{influxdb}, and the code changes indicate that this commit activates the costly profiling of Go by default.
Looking at the total study period, simple queries are improved by about $5\%$ by the end of the evaluation and group-by queries show a slight regression.
Overall, we detect performance changes for all request types in both study objects.

\subsection{Optimized Microbenchmark Suite\label{subsec:results-opt}}

\noindent
After generating the call graphs for both application benchmark and microbenchmark suite, we determine the practical relevance for both SUTs, and find an optimized microbenchmark suite based on the approach described in \cref{sec:background}, Removing Redundancies in Microbenchmark Suites. 

\free
\textbf{Computing Optimized Suites}

\noindent
For \victoria{}, the call graph analysis shows that $634$ project functions are called during the application benchmark of which $314$ are covered by microbenchmarks, thus indicating a practical relevance of $\approx 49\%$.
In the next step, by removing redundancies in the suite, the same relevance is already achieved with only $17$ microbenchmarks.
Many of these microbenchmarks, however, cover only a few additional nodes (three or less) of the application benchmark.
Thus, we use only the eight most relevant microbenchmarks for our further analysis, which corresponds to a $47\%$ practically relevant microbenchmark suite.

Initial experiments with the microbenchmark suite of \influx{} showed that there are major changes in the microbenchmark suite in the first $15$ commits, which also affects our potentially optimized microbenchmark suite:
a performance comparison of two versions of the microbenchmark is only possible if this microbenchmark is also present in both versions and has not been changed. 
Due to the fact that some of the most relevant microbenchmarks in the suite optimized for commit $1$ are missing in commit $15$, we set the base version to commit $15$ and shorten the evaluation period for the microbenchmarks.

The practical relevance of \influx{}'s complete microbenchmark suite is around $40\%$ at commit $15$ ($269$ of $660$ nodes overlap).
After computing the optimized suites without redundancies, many of the $26$ proposed microbenchmarks add only a few additional nodes to the overlap (three or less). 
Thus, similar to \victoria{}, we continue with only the $10$ most relevant microbenchmarks ($\approx 36\%$ practical relevance).

\begin{figure}[t]
	\centering
  \includegraphics[width=\linewidth]{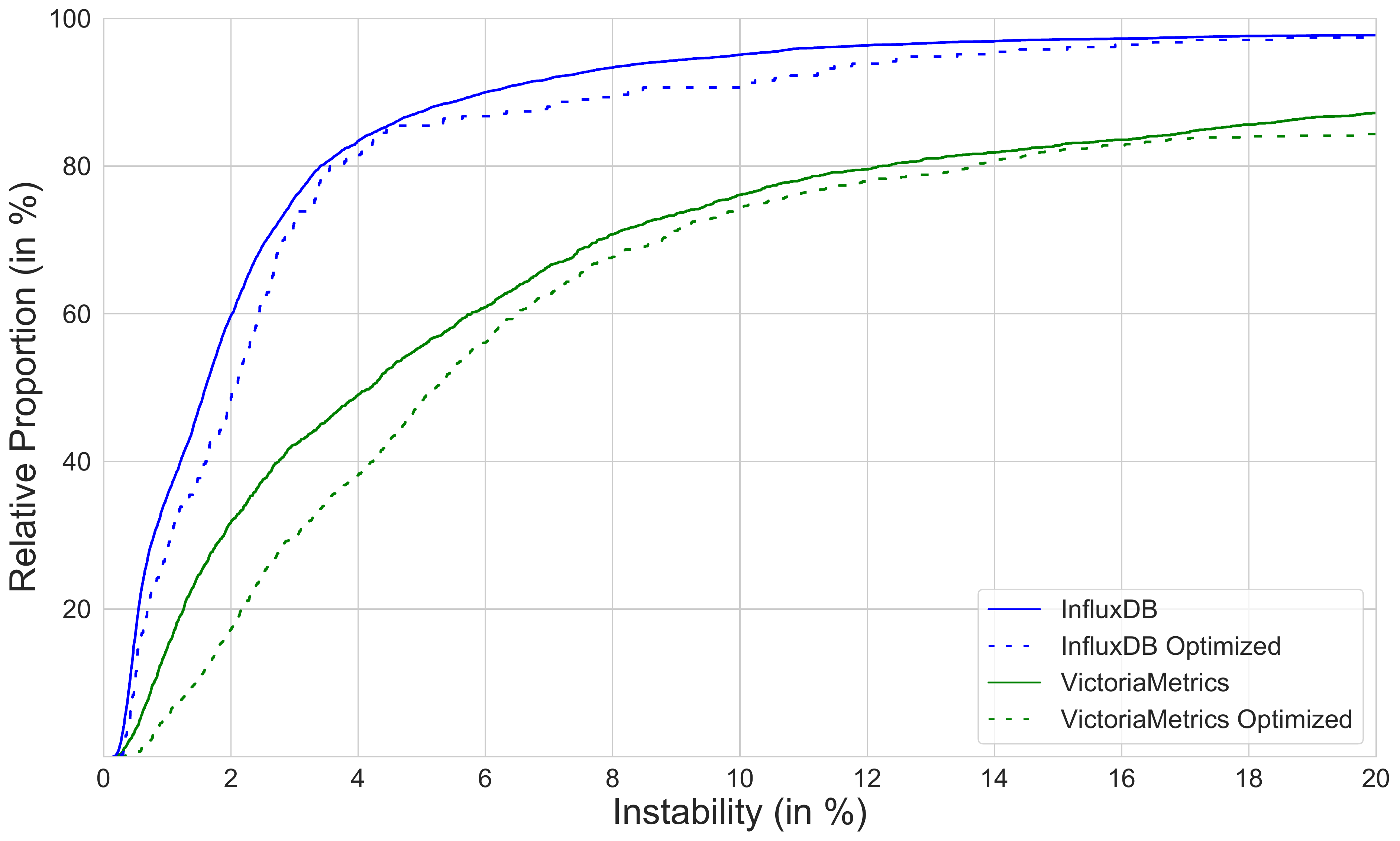}  
  \caption{\textbf{Microbenchmark instability.} Approx. $80\%$ of the microbenchmarks in the respective suites of \influx{} show an instability of less than $4\%$. For \victoria{}, however, only approx. $50\%$ of the microbenchmarks show an instability of less than $4\%$.}
  \label{fig:stability}
	\vspace{-1em}
\end{figure}

\free
\textbf{Determining Initial Detection Thresholds}

\noindent
\Cref{fig:stability} shows cumulative distribution functions for the instabilities of all microbenchmark suites in A/A experiments. 
While the microbenchmarks of \influx{} are very stable and $\approx 80\%$ of the measurements have an instability below $4\%$, the performance of \victoria{}'s microbenchmark suite(s) fluctuates more.
Here, only $\approx 50\%$ show an instability less than $4\%$.
Thus, to cover $\approx 80\%$ of the microbenchmarks instabilities in the respective suites with the initial detection threshold (see \cref{sub:analysis}, Jump and Trend Detection), we choose a general starting threshold of $12\%$ for \victoria{} and $6\%$ for \influx{} for our change detection, i.e., only experiments exceeding these thresholds in the first code changes are classified as performance changes.
In the subsequent code changes, this threshold adapts to the respective microbenchmark's instability and the algorithm will detect changes more reliably.

\free
\textbf{Detected Changes for \victoria{}}

\begin{figure}[t]
	\centering
  \includegraphics[width=\linewidth]{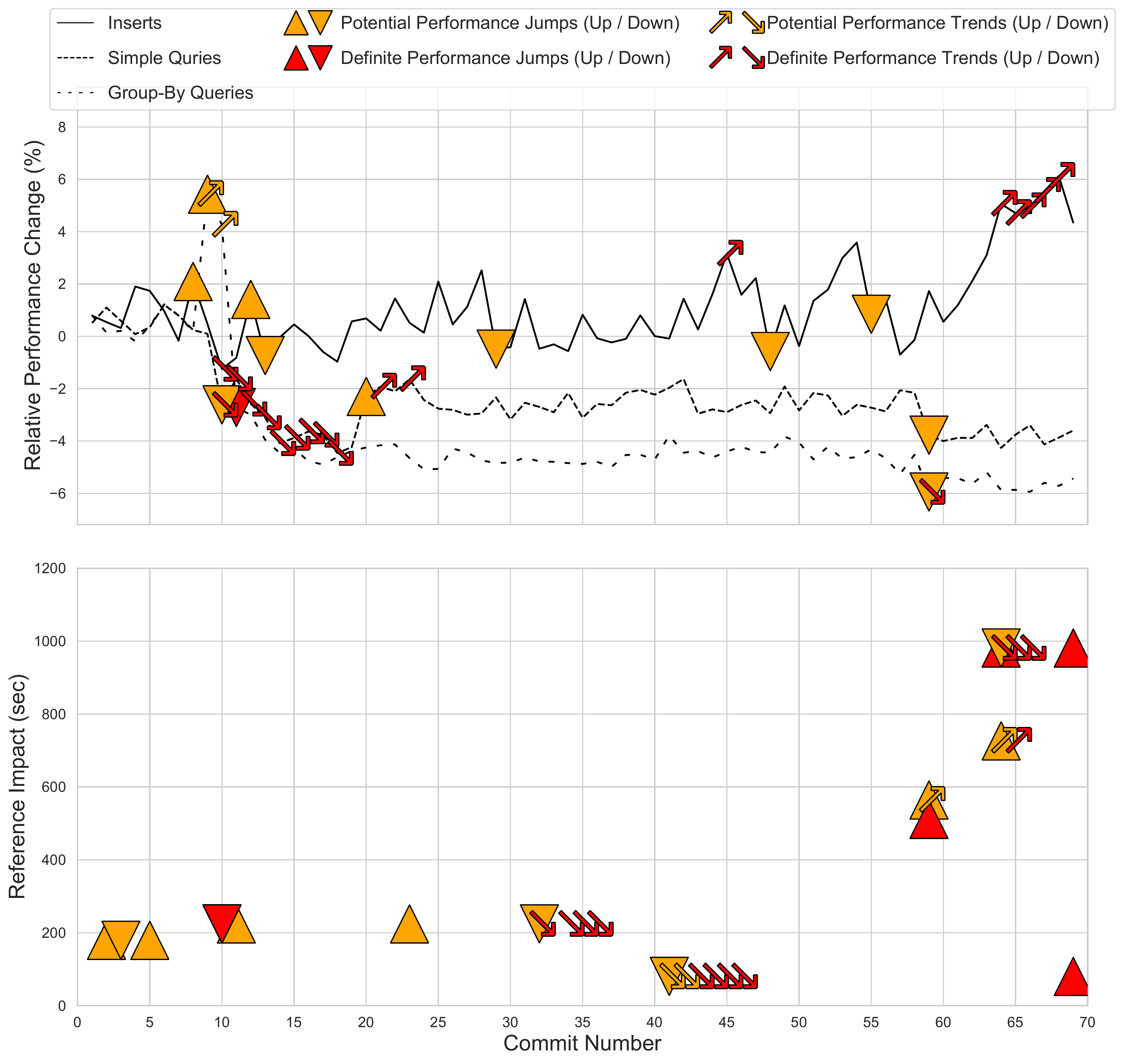}  
  \caption{\textbf{Results from the optimized suite for \victoria{}.} The upper part shows the results of the application benchmark and its detections while the lower part shows the detections from the microbenchmarks (higher means more likely impact on application performance). The optimized suite detects an insert-related change at commit $10$, a performance change for queries at commit $59$, and slower inserts at commit $64$ and $65$.}
  \label{fig:appAndMicroVm}
	\vspace{-1em}
\end{figure}

\begin{figure}[t]
	\centering
  \includegraphics[width=\linewidth]{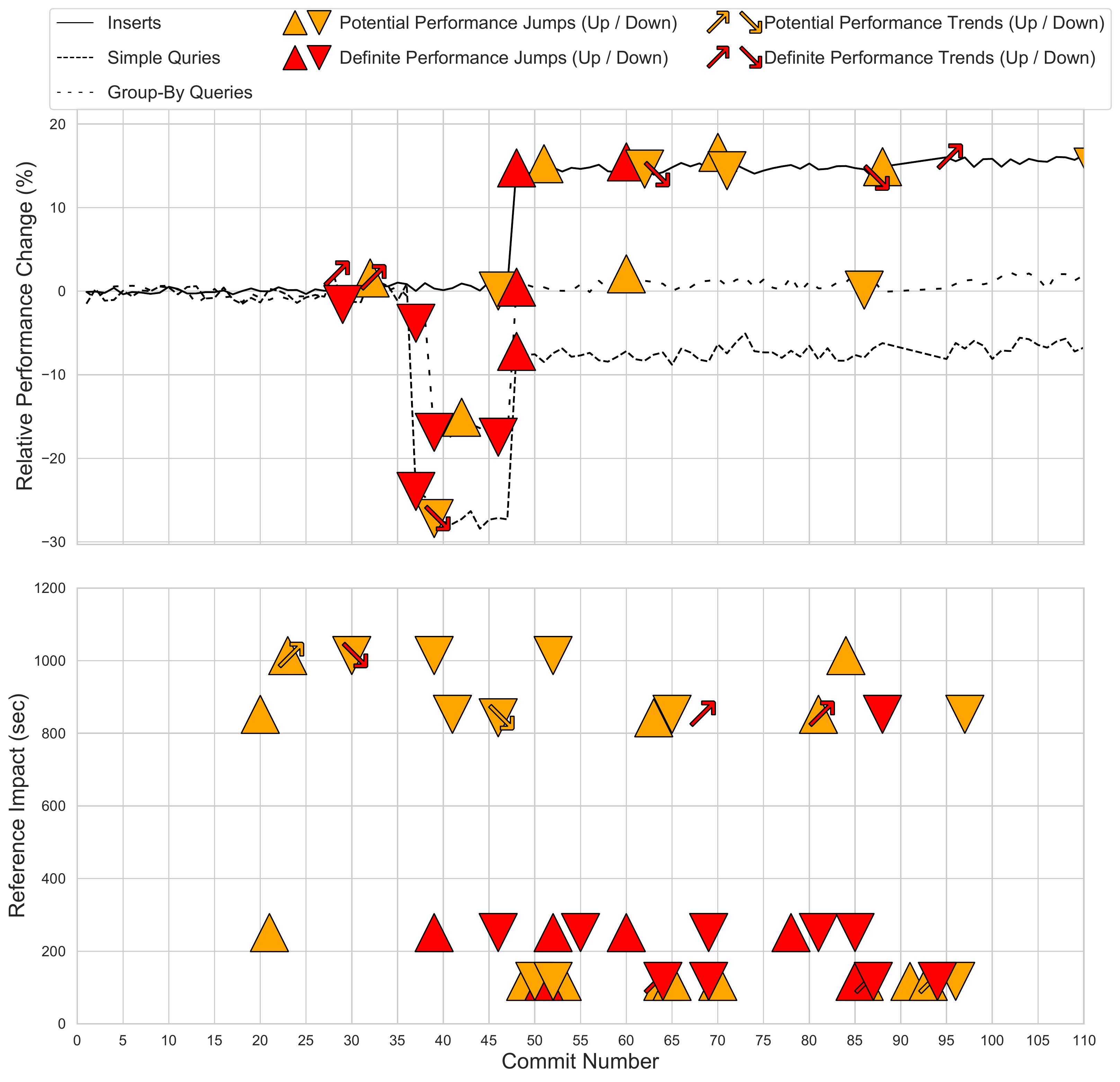}  
  \caption{\textbf{Results from the optimized suite for \influx{}.} The upper part shows the results of the application benchmark and its detections while the lower part shows the detections from the microbenchmarks (higher means more likely impact on application performance). The optimized suite of \influx{} with $\approx 36\%$ practically relevance detects five true positive alarms but also raises false alarms for $27$ commits.}
  \label{fig:appAndMicroInflux}
	\vspace{-1em}
\end{figure}

\noindent
Running both optimized microbenchmark suites detects multiple potential and definite performance changes for several microbenchmarks.
\Cref{fig:appAndMicroVm,fig:appAndMicroInflux} combine these detections with their corresponding reference impacts and evaluated performance metrics from the application benchmark.
Ideally, any relevant performance change in the application benchmark (line chart in the upper part of the figure) should also be detected by a microbenchmark with a large reference impact (lower part of the figure).
Nevertheless, because the microbenchmark suites in general only cover $47\%$ and $36\%$ of the application benchmark, we cannot expect to detect all changes.

\textbf{}

\free
Most of the detections in \victoria{}' optimized microbenchmark suite originate from microbenchmarks with a reference impact of about $200$ seconds in the application benchmark. 
With a total execution duration of the application benchmark of around $30$ minutes (without setup), their covered functions are responsible for approximately $10\%$ of the execution duration of the application benchmark. 

The first three potential jumps are false positives due to the moving dynamic threshold. 
At the beginning of the evaluation period, the dynamic threshold is not yet adjusted to the observed instability. 
Therefore, we do not consider them further. 

The first definite change in commit $10$ and the next potential jumps and definite trend until commit $36$ originate from the microbenchmark \textit{BenchmarkAddMulti}, evaluating \textquote{a fast set for uint64}~\cite{victoria} using buckets, which indicates a relevance for inserts. 
The change detection of the application benchmark, on the other hand, also identifies a definite trend and faster inserts at commit $10$.
Visual analysis shows that all further detections of the microbenchmark are not clearly reflected in the performance of the inserts at \victoria{}.

The detected changes from commit $41$ to $46$ refer to the microbenchmark \textit{BenchmarkRowsUnmarshal}, which evaluates the unmarshalling of the Influx line protocol (which we use in our benchmarking client). 
Similarly to the detections before, these are not reflected in the application benchmark's detected changes.

The next definitive change correlates with another potential trend and jump at commit $59$.
The corresponding microbenchmarks \textit{BenchmarkMergeBlockStreamsFourSourcesBestCase} and \textit{BenchmarkMergeBlockStreamsFourSourcesWorstCase} merge multiple block streams and are related to queries. 
Although the correlated benchmarks have a longer average execution time, both query types improve in the application benchmark at commit $59$ (we discuss this in \cref{sec:discussion}).

The microbenchmark change detection then raises signals at commit $64$ and $65$ for microbenchmarks related to insert requests. 
These significant signals with a reference impact of $981$ seconds (\textit{BenchmarkStorageAddRows}) and $726$ seconds (\textit{BenchmarkIndexDBAddTSIDs}) are also significantly noticeable in the application benchmark. 

Finally, the microbenchmarks \textit{BenchmarkRowsUnmarshal} and \textit{BenchmarkStorageAddRows} detect a definite change at commit $69$.
This change, however, is not visible in the application benchmark. 

In total, the optimized suite detects four true positives (commits $10$, $59$, $64$, and $65$), but also raises false alarms for $17$ commits.
On the other hand, the optimized suite does not detect the negative performance trend of group-by queries starting with commit $20$ (we discuss false negatives in more detail later).  

\textbf{}

\free
\textbf{Detected Changes for \influx{}}

\noindent
The optimized suite of \influx{} detects several potential and definite performance changes in five different microbenchmarks (see \cref{fig:appAndMicroInflux}). 

The microbenchmark with the largest reference impact, \textit{BenchmarkCreateIterator} ($1013s$ impact), accounts for around $40\%$ of the application benchmark's execution duration and benchmarks the creation of iterators for shard data items. 
This query-related benchmark identifies five potential performance changes for the commits $23$, $30$, $39$, $52$, and $84$.
While the commits $23$, $30$, and $52$ are configuration-related code changes, which are unlikely to have an impact on application performance, commit $39$ and $84$ introduce larger changes.
Commit $39$ updates a flux dependency and this improvement is also visible in the application benchmark for both query types. 
Commit $84$ adds a profiler option and modifies $68$ lines in the \textit{query.go} file.
The query performance in the application benchmark, however, is not affected by this change. 

The second most relevant microbenchmark is named \textit{BenchmarkWritePoints} ($942s$ impact) and it \textquote{benchmarks writing new series to a shard}~\cite{influxdb}, thus affecting insert requests. 
The first three detections are potential changes at commits $20$, $41$, and $65$, which introduce minor features or fix small bugs.
None of the three potential detected changes are visible in the application benchmark.
The detected definite trend at commit $68$ is caused by a minor configuration-related change.
Neither this one, nor the changes from the previous commits (the root cause for the trend detection might also be in earlier commits), however, show any performance change in application performance.
Next, commit $81$ introduces an optimization which is identified as a potential jump and a definite trend. 
This optimization, however, does not have any influence on the application performance. 
Commit $88$ is a minor change but also updates the flux dependency.
The application benchmark, on the other side, also detects a performance change for inserts.
Finally, there is a minor fix at commit $97$ identifying a potential change which is not relevant for application performance.

The third most relevant benchmark \textit{BenchmarkParsePointsTagsUnSorted} ($842s$ impact) benchmarks parsing of values and detects potential changes at commit $46$ and $63$ which are both also detected by the application benchmark.
Commit $46$ changes a default parsing option and this is also reflected in a potential improvement signal for inserts and a definite one for group-by queries in the application benchmark. 
The change introduced with commit $63$ prevents a formatting of time strings in certain situations.
This improvement is also detected as a potential improvement for inserts in the application benchmark. 

The next microbenchmarks, \textit{BenchmarkDecodeFloatArrayBlock} ($251$ seconds) and \textit{BenchmarkIntegerArrayDecodeAllPackedSimple} ($116$ seconds), decode array blocks of float64 and integer values and have a significantly lower reference impact.
The float benchmark detects one potential and nine definite changes, but only the changes at commit $39$ (already identified by the most relevant microbenchmark), $46$ (already identified by the third most relevant microbenchmark), and $60$ are also detected by the application benchmark.
Commit $60$ fixes a cache-related race condition and this also impacts the performance of inserts and simple queries in the application benchmark. 
All other detections, however, are false positives. 
The least relevant integer benchmark detects changes for $17$ commits. 
Here, five of the $17$ detections (for commits $51$, $63$, $70$, $87$, and $94$) correspond to the detections of the application benchmark for inserts and one matches a detection for grouping queries (commit $86$).
Nevertheless, because all changes introduce only minor features and smaller bug fixes, which are not related to any core functionality, we assume no direct correlation and consider all of them as false positives.

In total, the optimized suite detects five true positives (commits $39$, $46$, $60$, $63$, and $88$), but also raises false alarms for $27$ commits.
Moreover, the optimized suite could not detect the two major performance changes at commit $37$ and $48$.
While the performance change at commit $37$ might not be detected because there is no microbenchmark covering the relevant code sections, the change at commit $48$ can not be detected because it is related to the runtime environment. 

\subsection{Complete Microbenchmark Suite\label{subsec:results-micro}}

\begin{figure}[t]
	\centering
  \includegraphics[width=\linewidth]{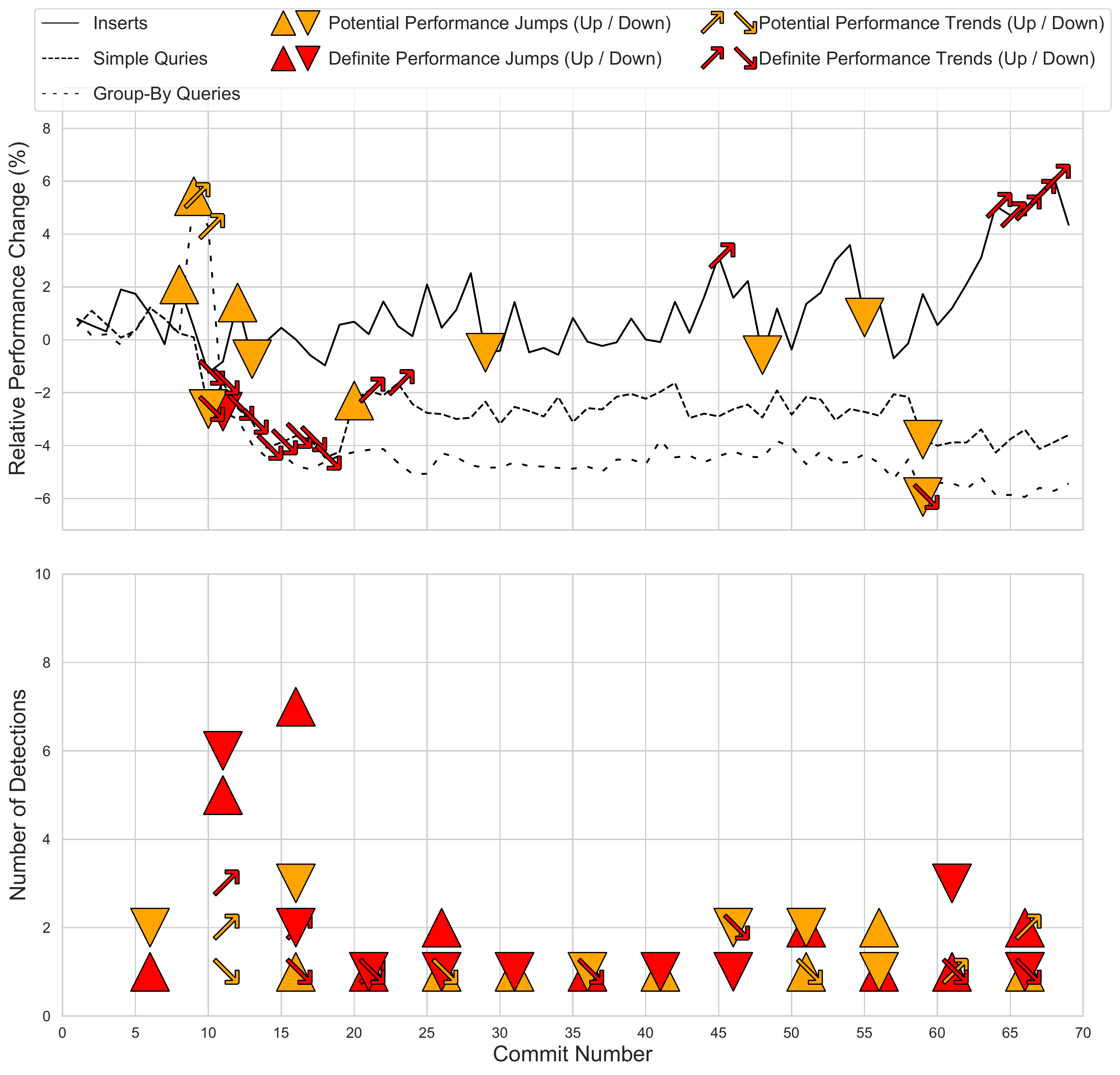}  
  \caption{\textbf{Complete suite results for \victoria{}.} The complete suite with $177$ microbenchmarks in total detects $91$ changes that can not be directly linked to the application-relevant performance metrics.}
  \label{fig:appAndMicroFullVm}
	\vspace{-1em}
\end{figure}

\begin{figure}[t]
	\centering
  \includegraphics[width=\linewidth]{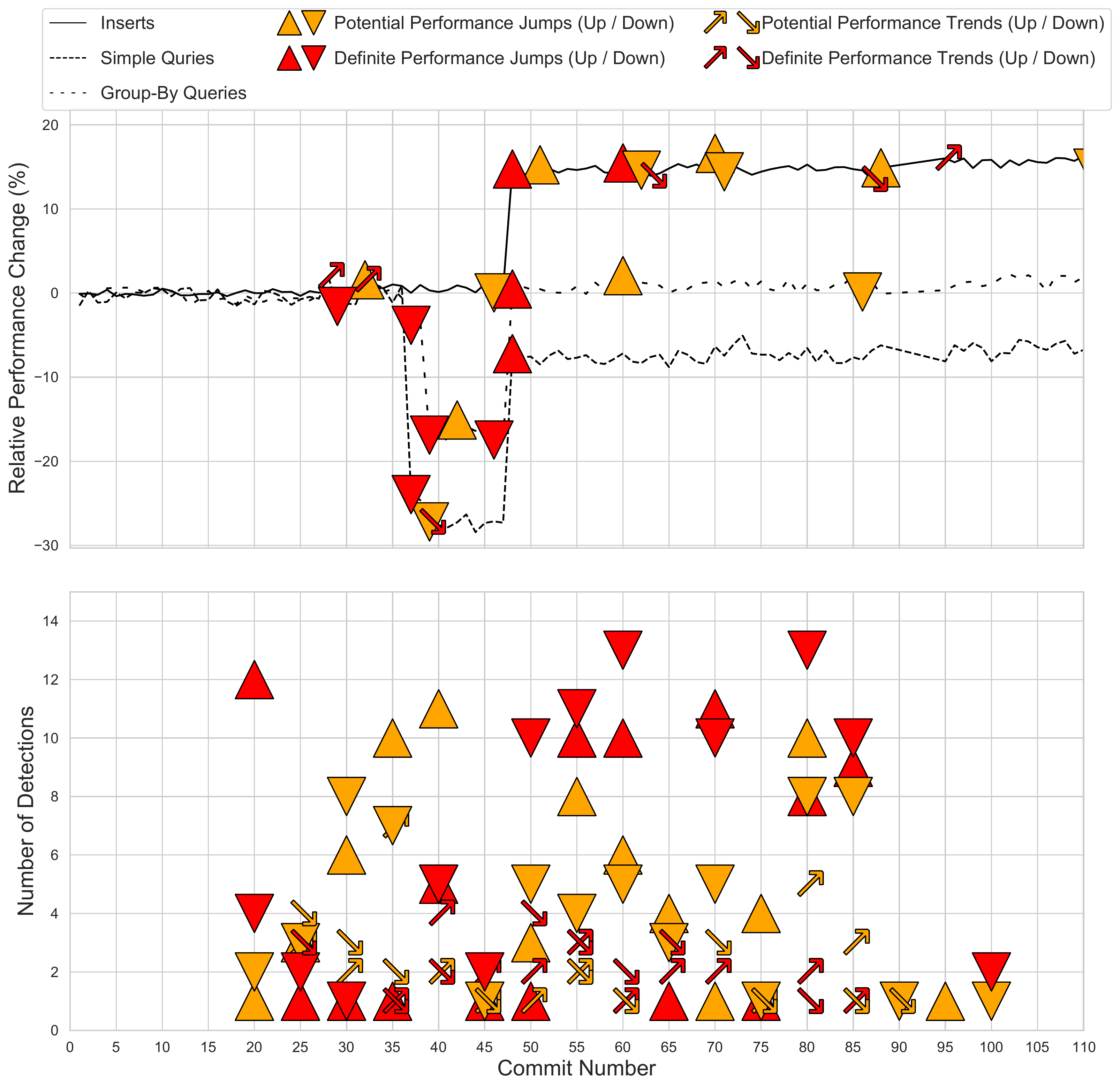}  
  \caption{\textbf{Complete suite results for \influx{}.} The suite shrinks down from $426$ to $109$ microbenchmarks during the evaluation period. Nevertheless, the complete suite detects $392$ performance changes that can not be mapped to the application-relevant metrics.}
  \label{fig:appAndMicroFullInflux}
	\vspace{-1em}
\end{figure}

Running the complete suite with hundreds of microbenchmarks for each commit in practice is unrealistic, as it is too expensive and time-consuming to do so.
Nevertheless, to rate the improvement and better compare the optimization technique to this alternative, we execute the complete suite for every fifth commit.
In contrast to the optimized suite, we cannot use the reference impact to rank the results because many microbenchmarks do not overlap or only barely overlap with the reference application benchmark call graph. 
Thus, we aggregate the respective detection when interpreting the results, e.g., if ten microbenchmarks detect a definite change, this change might be practically relevant. 
Moreover, we also adapt the dynamic detection threshold to the evaluation of every fifth commit only and consider only the last three values (instead of $10$).

\Cref{fig:appAndMicroFullVm,fig:appAndMicroFullInflux} show the application metrics (above) and the detections from the complete microbenchmark suite (below).

\free
\victoria{}'s complete microbenchmark suite detects $91$ changes in total and those cover all evaluated code changes.
In particular, we observe that there are not only multiple detections for each commit, but that these detections are also often contradictory (\citet{chen_exploratory_2017} also report this phenomenon).
Except for commit $46$, where all detected changes are improvements, there are always at least one microbenchmark each measuring a performance degradation and improvement respectively.

The complete suite of \influx{} detects $392$ performance changes in total. 
Similarly, the suite detects contradictory performance changes at each commit and these cannot be matched with the metrics of the application benchmark.

In total, the complete suites detect hundreds of performance changes at high cost but only some of them are relevant. 
Without further information and criteria, such as an impact or relevance factor, it is impossible to identify the relevant ones.

\subsection{Findings and Implications\label{subsec:results-impl}}

In total, we examined $180$ code changes in two open-source TSDBs using $540$ application benchmark runs, $495$ executions of optimized microbenchmark suites, and $102$ runs of complete microbenchmark suites.
These correspond to approximately $1,900$ hours of benchmark execution duration.
Despite this vast number of experiments, we cannot demonstrate a clear benefit of using an optimized microbenchmark suite: while some benefits exist, there are limitations.
Overall, our results help to better understand the trade-off between the execution of application benchmarks, optimized, and complete microbenchmark suites (see \cref{tbl:costs,tbl:changesApp}). 

\free
\textbf{Application Benchmarks}

\noindent
The setup of an automated application benchmark is complex and time-consuming.
It requires scripts for starting the SUT and client instances, triggering and orchestrating the benchmark, collecting the measurements, and finally for analyzing the measurements.
Running this complete pipeline took about $40min$ for \victoria{} and $130min$ for \influx{} in our experiments, which corresponds to costs of about $\$0.13$ and $\$0.40$ per experiment repetition.
Once set up, however, an application benchmark is a great tool for reliably detecting performance regressions or improvements in code changes.
In our studied systems, this advantage can be illustrated especially with \influx{}: 
A clear improvement caused by a new feature and a clear drop caused by a misconfiguration (which is impossible to detect using a microbenchmark) can be directly linked to specific commits.  
Furthermore, although small performance shifts between two successive commits may not be detected due to variability, an application benchmark can also be used to reliably detect performance trends. 
We can observe this characteristic especially for \victoria{}: 
both query types show performance improvements between commit $10$ and $20$, but due to the large confidence intervals the changes cannot be directly connected to a single commit.
Finally, our experiments also show that the Duet Benchmarking technique can not be applied everywhere without further modifications.
Due to a non-deterministic characteristic of \victoria{}, it is difficult to evaluate insert operations accurately.

Our experiments show that a well-designed application benchmark can reliably detect performance changes even in highly variable cloud environments.
In our use cases, an application benchmark is relatively fast and cost-efficient, because we have chosen a rather simple setup with only two instances.
In more complex setups using more complex application benchmarks, however, the price per benchmark will be higher, and the execution may also take longer. 
These more complex benchmarks include different load scenarios, involve many more instances and components, or evaluate the impact of changes in the environment, e.g., network fluctuations or (temporal) outages of individual components~\cite{silva_cloudbench_2013, hasenburg_mockfog_2021}.
Thus, depending on the frequency of code changes, we argue that an application benchmark should usually be scheduled to run daily, weekly, or after major code changes. 

\begin{table}[t]
\centering
\caption{\textbf{Benchmarking durations and prices.} Running complete microbenchmarks suites takes a lot of time while an optimized suite is faster and less expensive than an application benchmark.}
\begin{adjustbox}{max width=\columnwidth}
\begin{tabular}{@{}llrr@{}}\toprule
	Benchmark &  & \victoria{} & \influx{} \\
	\midrule
	Application Benchmark &  & $\sim 40min$ & $\sim 130min$\\
	 &  & ($\sim \$0.13$) & ($\sim \$0.40$)\\
	\\
	Optimized Suite &  & $\sim 20min$ & $\sim 40min$\\
	 &  & ($\sim \$0.03$) & ($\sim \$0.06$)\\
	\\
	Complete Suite &  & $\sim 4h$ &  $\sim 11h$\\
	 &  & ($\sim \$0.38$) &  ($\sim \$1.05$)\\
\bottomrule
\end{tabular}
\end{adjustbox}
\label{tbl:costs}
\vspace{-1em}
\end{table}

\begin{table}[t]
\centering
\caption{\textbf{Result summary.} While application benchmark (app) and optimized microbenchmark suite (opti) detect a rather small number of performance changes, the complete suite (full) finds a lot more.}
\begin{adjustbox}{max width=\columnwidth}
\begin{tabular}{@{}lrrrrrr@{}}\toprule
	& \multicolumn{6}{c}{Number of definite changes (+potential)} \\
	Project & \multicolumn{3}{c}{\victoria{}} & \multicolumn{3}{c}{\influx{}} \\
	Benchmark & \multicolumn{1}{c}{App} & \multicolumn{1}{c}{Opti} & \multicolumn{1}{c}{Full} & \multicolumn{1}{c}{App} & \multicolumn{1}{c}{Opti} & \multicolumn{1}{c}{Full}\\
	\midrule
	Jump up & $0 (+4)$ & $5 (+6)$ & $23 (+14)$ & $4 (+6)$ & $6 (+14)$ & $72 (+77)$\\
	Jump down & $1 (+5)$ & $4 (+4)$ & $17 (+15)$ & $5 (+6)$ & $10 (+10)$ & $83 (+67)$\\
	\midrule
	Trend up & $8 (+2)$ & $1 (+2)$ & $6 (+5)$ & $3 (+0)$ & $4 (+2)$ & $20 (+30)$\\
	Trend down & $11 (+0)$ & $11 (+2)$ & $7 (+4)$ & $3 (+0)$ & $1 (+1)$ & $19 (+24)$\\
\bottomrule
\end{tabular}
\end{adjustbox}
\label{tbl:changesApp}
\vspace{-1em}
\end{table}

\free
\textbf{Complete Microbenchmark Suite}

\noindent
Running the complete microbenchmark suites of our evaluated projects takes around $4h$ for \victoria{} and $11h$ for \influx{}.
Thus, evaluating one commit using one single experiment costs about $\$0.38$ for \victoria{} and about $\$1.05$ for \influx{}. 
For reliable measurement results, this single experiment should be run at least 3 times (concurrently), thus multiplying the cost.
Furthermore, if a microbenchmark detects a performance change, the exact evaluation of the results is still hard due to the large number of experiments, instability of microbenchmarks, and it is often unclear to which degree a change affects the production environment and application-relevant metrics.
For example, when running the complete suites for every fifth commit, we observe hundreds of performance changes in the microbenchmarks (see \cref{tbl:changesApp}), but these are not reflected in the application-relevant benchmark metrics.

Running and evaluating a complete microbenchmark suite is usually expensive, takes a long time, is difficult to evaluate, and hardly yields any findings or findings that are difficult to derive.
If code changes happen at intervals of minutes or hours, then this type of benchmark is only suitable for nightly or weekly performance evaluations.
Nevertheless, a complete run can help to analyze a detected performance problem in more detail, help to isolate the issue, and find the root cause.
Hence, it could be triggered whenever an application benchmark run has identified a performance change.  

\free
\textbf{Optimized Microbenchmark Suite}

\noindent
The optimized microbenchmark suite without redundancies runs much faster (around $20min$ and $\$0.04$ for \victoria{}; around $45min$ and $\$0.07$ for \influx{}), is easier to evaluate and, if covering practically relevant parts, can detect the same performance changes that can also be detected by an application benchmark (see \cref{tbl:changesApp}). 
Using optimized microbenchmark suites, we can identify four true positive detections for \victoria{}) and five true positives for \influx{}.
Nevertheless, both optimized suites also raise false alarms, especially through microbenchmarks with a low reference impact (which is part of the reason that the full suite detects so many false positives).

Running only practically relevant microbenchmarks significantly reduces the execution duration and also simplifies the analysis of the results.
If the optimized suite covers a large portion of the practically relevant code sections, the suite can quickly detect performance changes and link them to specific commits at low cost.
On the other hand, if performance changes relate to the runtime environment, integration, and or interaction of different application components, the microbenchmark suite cannot detect them.
The profiling setting, which caused a significant performance drop in \influx{} at commit $48$, can not be found in the microbenchmarks because it was caused by a general configuration in the production(-like) environment.
Another problem when using the microbenchmark suite are the benchmarks within the suite itself.
If the suite changes often and especially if this involves the most relevant microbenchmarks with a large reference impact, then a continuous comparison is not possible and the optimized suite has to be re-determined periodically. 
In our experiments, this problem affects \influx{} twice: 
once at the beginning of the evaluation period (commit $15$); and once at commit $80$.
Thus, while an optimized benchmark suite can evaluate the performance several times a day, this benchmarking strategy should not be the only benchmarking approach used.

\section{Discussion}
\label{sec:discussion}

Our experiments show that optimized microbenchmark suites can detect application performance changes in certain situations.
While both micro- and application benchmarks may not be suitable for more detailed analysis of every commit in large projects with many code changes, they are still suitable for daily (or nightly) and weekly use as well as for a more detailed analysis after major changes.
An optimized microbenchmark suite covering large practically relevant code parts can complement this by providing a fast performance feedback.
Nevertheless, there are some limitations and possible extensions which we discuss in the following.

\free
\textbf{The Trade-off Between Cost and Accuracy}

\noindent
Within our experiments, we can produce reliable and reproducible results with three experiment repetitions. 
Nevertheless, several microbenchmarks show wide confidence intervals of more than $20\%$ and are unstable. 
For each project, it is thus essential to find a good compromise between effort and cost on one side and accuracy and reliability on the other side.

Besides narrowing the CIs through additional experiment repetitions, which also increases cost accordingly, there are further optimizations by stopping benchmark runs under certain conditions or predicting unstable ones~\citep{alghmadi_automated_2016,alghamdi_towards_2020,he_statistics-based_2019,laaber_dynamically_2020,laaber_predicting_2021}. 
For example, stopping benchmarks as soon as there is a reliable finding might shorten the benchmark duration, excluding unstable microbenchmarks might avoid unnecessary effort, or multiple smaller microbenchmarks might be more reliable and thus more cost-effective than a large unstable one. 
In our study, excluding a large unstable microbenchmark would just reduce the practical relevance by removing some microbenchmarks from the optimized suite of both study objects without adding equally relevant ones.
Thus, this is subject to further research.

\free
\textbf{Changes in the Optimized Suite over Time}

\noindent
In our study, we use a fixed code state to optimize the microbenchmark suite, i.e., commit No. $0$ for \victoria{} and commit No. $15$ for \influx{}.
This base version should not be changed as long as possible to generate a long measurement series for trend detection.
Nevertheless, there are situations in both application benchmarks and microbenchmarks where this base version has to be reset and the optimization has to be repeated. 
Thus, the optimized suite cannot be considered static and has to be changed from time to time.

Both types of benchmark require a new base version when the benchmark itself is modified. 
Regarding the application benchmark this is, e.g., the case if the workload is no longer realistic and needs to be adjusted (e.g., the number of customers has doubled, which means twice as many requests in the production system). 
An adjusted application benchmark will imply a changed reference call graph and updated reference impact values.

Regarding the microbenchmarks, for example, there is the modification of the invocation parameters and that individual microbenchmarks might be removed (as can be seen in our experiments) or new ones might be implemented.
Moreover, as every commit modifies the code that is evaluated by the microbenchmarks, the respective microbenchmark call graphs has to be updated as well. 
Both changes may require changes to the optimized suite as well.

\free
\textbf{Identifying False Alarms}

\noindent
In our experiments, both optimized suites detect nine true positive performance changes but also raise false alarms for $44$ commits in total.
These false alarms are caused, among other things, by measurement inaccuracies, but can also be caused by the approach reacting to changes in non-practically relevant functions.
For example, if the performance of a non-practically relevant function degrades, but this function is also evaluated by a microbenchmark with large reference impact (e.g., the uncovered function of MB1 in \cref{fig:approach}), then the performance of the microbenchmark will also degrade, even though this change has no impact on application performance.
Confirming a detected change or spotting a false alarm would require to start an application benchmark in a realistic setup, which would imply corresponding costs.  
Thus, identifying false alarms in advance would be major improvement in further research. 

Besides using the reference impact as additional classification for the reliability of detections, for example, the detected changes could be flagged and stored if the respective microbenchmark raised a false alarm.
Using this history of changes, it might be possible to determine a reliability value for each microbenchmark which can be used to assess whether their detected performance change should be disregarded or not. 
A microbenchmark that successfully detected application performance changes in the past might also do this for future code changes.

Moreover, tagging microbenchmarks that are affected by a code change (i.e., only a fraction of the optimized suite), might also ease the result analysis. 
If one of those microbenchmarks raises an alarm, it is worth a detailed evaluation because there is a related code change. 
If a detected change is not raised by a tagged microbenchmark, it might be a false alarm.
Furthermore, it might even be feasible to run only those microbenchmarks that cover modified functions. 

\free
\textbf{Interpretation of Microbenchmark Performance Changes}

\noindent
To derive concrete implications from statements such as ``microbenchmark A's performance has dropped by $5\%$'', there are several options.
In the optimal case, application developers know the underlying logic of the respective microbenchmark, can directly relate a detected change to the target functionality (e.g., the request type), and rate the impact on the production environment.
This is, however, not realistic, especially for large projects.
We suggest interpreting and storing the detected changes as warnings which will trigger an application benchmark to verify the overall system performance and or to use them to support root cause analysis when a future application benchmark shows significant performance changes and the originating code change needs to be identified. 

A strict policy that, for example, rejects a commit when a performance issue is detected by a microbenchmark would in many cases be incorrect.
In our experiments, for example, a microbenchmark detected that the merging of block streams takes longer for \victoria{} in commit $59$ while the application benchmark observed faster queries. 
Because this corresponding code update merges $8$ new features and fixes $6$ bugs, we can not identify the exact reason for this phenomenon due to its complexity, but we can find two possible explanations.
First, even though merging the block streams takes longer because more data is processed, the query latency decreases because fewer streams need to be merged, thus resulting in fewer calls to the respective function while running the application benchmark.
Second, while merging streams takes longer, another feature is introduced that improves the query latency but is not covered by the optimized microbenchmark suite (yet).
In such a scenario, the feature leading to the improvement might be the dominant code change while the microbenchmarks can only detect the less relevant degradation covered by the suite.

\free
\textbf{Implications for Production}

\noindent
Our extensive experiments using two time series database systems show many interesting aspects when running optimized microbenchmark suites.
Nevertheless, there is no general (micro-) benchmarking strategy that can simply be applied to every project. 
The strategy needs to be determined individually for each project and depends, among other things, on the general development progress, the number of code changes per day, the production environment, the expected load, and the impact of a potential performance issue.

Optimized microbenchmark suites can be a helpful tool for large projects with multiple developers, a large code base, and many code changes per day (e.g., our studied time series database systems).
Here, a detailed performance evaluation of every code change is not possible, but optimized suites can help to evaluate these changes well enough, i.e., covering practically used code sections. 
In smaller projects it can also be useful to save costs. 
For example, a cost-intensive execution of application benchmarks for each code change possibly can be replaced with the optimized suite while the application benchmark is, e.g.,  executed only weekly, for every 10th commit, or for each major change.

As the concrete parameter values have to be defined individually for each project, we recommend analyzing past code changes and running some trial benchmarks first, e.g., to estimate variances.
Based on these results, it is then possible to derive concrete values such as (micro-) benchmark frequency, detection thresholds, or actions in case of a detected performance change.

\free
\textbf{Limitations of Application Benchmarks}

\noindent
Application benchmarking offers the possibility of placing the evaluated system in any requested situation.
From examining increased usage during holiday season to studying the effects of component failure, application benchmarks can be implemented for many situations.
Nevertheless, they effectively use an artificial load and do not run on the production system, which also has limitations.
For example, the actual production load may not match the load assumed by the benchmark, resulting in different results and implications.
In addition, not all use cases can provide a second environment that can be used for benchmarks.
For example in IoT scenarios, it is hardly possible to maintain a second identical building with the same smart home devices just for testing and benchmarking purposes. 
Alternatively and or complementary to application benchmarks, among others, application performance monitoring, gradual roll-outs, or dark launches can be used to detect performance changes in production.\footnote{If it is possible to record call graphs in the production environment, these graphs can also be used as (a real) reference to compute the optimized suite.}
While benchmarking is used before deploying to production and does not affect real users, live testing techniques such as gradual roll-outs are applied in the real production environment.
Ideally, there should be a holistic combination of approaches from both phases, before and during live deployment.

\free
\textbf{Further Improvements and Research Directions}

\noindent
In our experiments, we benchmark successive code changes in the commit history of two large open-source TSDBs and analyzed them in detail.
Nevertheless, our findings can not be generalized to all systems.
There might be combinations of microbenchmark suites, SUTs, application benchmarks and their evolution over time in which microbenchmarks can detect performance regressions with neither false positives nor false negatives.
We believe that our findings are representative for most real world combinations.
This is based on the intuition that microbenchmark suites are unlikely to always have full code coverage of the SUT and that the functions studied by individual microbenchmarks may or may not have significant effects on the execution duration of application benchmarks.
Overall, our study motivates further research on the computation, usage, and advantages of optimized microbenchmark suites.
\section{Related Work}
\label{sec:related}

There is extensive previous work on benchmarking of software systems, e.g., in the context of cloud storage systems~\cite{bermbach_benchfoundry_2017, cooper_benchmarking_2010, muller_benchmarking_2014, paper_difallah_oltpbench,paper_pallas_security_performance_hbase}, which provides the application benchmarks necessary for this work. 
Regardless of the type of SUT and kind of benchmark, all benchmarks should aim for design goals such as relevance, portability, or repeatability to provide reliable results~\cite{jiang_survey_2015, paper_bermbach_benchmarking_middleware, folkerts_benchmarking_2013, huppler_art_2009, bermbach_cloud_2017}.
In the following, we discuss related work to our study focusing on benchmarking in CI/CD pipelines, approaches to reducing the overall benchmark execution time, dealing with cloud variability, and approaches for detecting and quantifying performance changes.
To the best of our knowledge, we are the first who use optimized microbenchmark suites as proxy for application benchmarks.

\free
\textbf{Benchmarking in CI/CD Pipelines}

\noindent
The idea of using performance testing or benchmarking in CI/CD pipelines, also in cloud environments, has already been addressed in several related papers.
\citet{mostafa_tracking_2009} argue that application performance after new commits should be tracked and propose an automatic approach based on call trees. 
\citet{foo_mining_2010} manually inject performance issues in three study objects to verify their automatic performance regression detection approach~\cite{foo_mining_2010, foo_industrial_2015}. 
\citet{waller_including_2015} include microbenchmarks in a CI/CD pipeline.
Moreover, several studies also conduct performance case studies using a dedicated benchmarking step and real software projects~\cite{grambow_continuous_2019, daly_industry_2019, ingo_automated_2020, daly_creating_2021}.
\citet{javed_perfci_2020} propose a CI/CD tool chain considering performance tests.
\citet{silva_cloudbench_2013} and \citet{hasenburg_mockfog_2021} propose frameworks supporting the automatic execution of benchmark experiments in the cloud as part of CI/CD pipelines. 

Our paper continues this research by studying, through extensive experimentation, to which degree different benchmarking approaches can detect performance changes of open source systems as part of a CI/CD pipeline. 

\vspace{1em}
\free
\textbf{Reducing Overall Benchmark Execution Time}

\noindent
One challenge in benchmarking is the high execution duration of benchmarks and the resulting costs.
Aside from the optimization strategy for microbenchmark suites that we use in this paper~\cite{grambow_using_2021}, there are other approaches towards reducing the execution duration. 

Test case prioritization usually focuses on functional unit test~\cite{rothermel_test_1999}, but can also be applied to microbenchmarks.
\citet{mostafa_perfranker_2017} prioritize test cases in performance regression testing,
\citet{laaber_applying_2021} apply several test case prioritization techniques to a number of microbenchmark suites.
One key finding from this study is that the top three major performance changes can be identified after running $29\%$ to $66\%$ of the complete microbenchmark suite, which demonstrates the potential of optimizing microbenchmark suites.

\citet{de_oliveira_perphecy_2017} propose another approach for reducing the benchmark suite by analyzing the SUT binary to select microbenchmarks based on code change indicators and information from prior benchmark runs.
This approach, however, treats every code section equally and does not favor practically relevant code, i.e., functions that are actually used in production.

\citet{chen_perfjit_2020} use the functional tests of software projects and extract classifiers for predicting tests that will reveal performance changes. 
Overall, this reduces the testing time drastically and the approach is also able to detect real performance issues in production.
Nevertheless, the authors use a partly automatic, partly manual performance analysis based on reported issues for their SUTs only and the number of application-relevant performance changes thus might be underestimated. 

Finally, there are several approaches that stop benchmark runs when the SUT is stable and is unlikely to produce different results with more load or repetitions~\cite{alghmadi_automated_2016,alghamdi_towards_2020,he_statistics-based_2019,laaber_dynamically_2020}.

In this paper, we followed the optimization strategy from our previous work~\cite{grambow_using_2021}, because it combines application and microbenchmarks.
This can be combined or replaced with other microbenchmark prioritization strategies to reduce the number of false alarms (e.g., ~\cite{mostafa_perfranker_2017, laaber_applying_2021}) or to further shorten the execution duration by stopping microbenchmarks once the results are stable and or do not show significant performance changes (e.g., \cite{alghmadi_automated_2016,alghamdi_towards_2020,he_statistics-based_2019,laaber_dynamically_2020}).

\free
\textbf{Cloud Variability and Unstable Microbenchmarks}

\noindent
A key requirement of benchmarks, the repeatability, is difficult to realize in variable cloud environments due to the many random factors that affect the benchmark~\cite{folkerts_benchmarking_2013, silva_cloudbench_2013, bulej_duet_2020, binnig_how_2009, bulej_initial_2019, laaber_performance_2018, laaber_evaluation_2018, schad_runtime_2010, leitner_exploratory_2017, paper_difallah_oltpbench, iosup_performance_2011, leitner_patterns_2016, uta_is_2020, paper_bermbach_expect_the_unexpected}. 
One way to minimize the effects of this variability and the number of experiment repetition is to benchmark multiple SUTs concurrently on the same VM(s)~\cite{bulej_initial_2019, bulej_duet_2020}.
This is, however, not always easy to implement or even possible.
In more complex systems that include several components distributed on different instances, for example, it would be necessary to ensure that the individual components are also exposed to the same load concurrently to provide a valid application benchmark.
To the best of our knowledge, we are the first who use and apply the Duet Benchmarking technique proposed by \citet{bulej_initial_2019} in longer running application benchmarks to counteract random cloud variability and to provide repeatable results. 

\citet{laaber_evaluation_2018} study several microbenchmark suites on bare metal and in cloud environments. 
Two key finding in their study are that not all microbenchmarks can be used to detect performance changes reliably due to large instability and that microbenchmark suites often contain high levels of redundancy. 
Our study confirms both.
The optimization technique eliminating redundancies reduces the number of microbenchmarks from $177$ to $17$ for \victoria{} and from $426$ to $26$ for \influx{}.
Moreover, both our study objects, and especially the suite of \victoria{}, contain microbenchmarks with an instability of more than $10\%$ (see \cref{fig:stability}).
To counteract this, \citet{laaber_predicting_2021} use machine learning to predict the stability of microbenchmarks without executing them. 
This approach could be combined with the optimization approach we followed in this paper to automatically exclude unstable microbenchmarks from the optimized suite or to prefer stable ones.

\free
\textbf{Detecting and Quantifying Performance Changes}

\noindent
Besides basic threshold metrics such as the ones we adapted from \citet{grambow_continuous_2019} by using a dynamic threshold which adjusts to the individual benchmark instability, there are more complex techniques for detecting and quantifying performance changes. 
\citet{foo_mining_2010} use performance signatures of past experiment runs and determine confidence measures. 
\citet{daly_industry_2019} also consider noise in their performance evaluation and cluster the time series experiment data to identify performance change points~\cite{matteson_nonparametric_2014}.
Moreover, even though the Iter8 framework proposed by \citet{toslali_iter8_2021} is designed for live testing, the proposed decision bayesian learning based algorithms can also be adapted to decide which version performs better. 
Finally, there are approaches that focus on automatically identifying the respective root causes of performance changes~\cite{nguyen_industrial_2014, heger_automated_2013}.
Each of these approaches could be used as alternatives for detecting performance changes and might, e.g., reduce the number of false alarms.
On the other hand, however, each of these approaches also increases the complexity and implementation effort of the analysis.

To the best of our knowledge, we are the first who apply a dynamically adapted performance detection threshold which adjusts along with the analyzed code changes to the respective micro or application benchmark instability. 

\section{Conclusion}
\label{sec:conclusion}

Both microbenchmarks and application benchmarks can be used in CI/CD pipelines to ensure that performance and non-functional requirements of software systems are met in every release. 
For large and complex projects with multiple code changes per day, however, both are too costly to examine every single code change in detail.

In this paper, we explored to which degree application-relevant performance changes, such as an increase in query latency, can also be detected by optimized microbenchmark suites.
For this, we use the commit history of \influx{} and \victoria{} and study them by running extensive benchmark experiments with application benchmarks, using an application coverage-based optimization strategy for microbenchmark suites, and running complete suites, we could show that this is indeed possible with some limitations.
As we discovered, the approach requires that existing microbenchmarks cover (almost) all application-relevant code sections but still results in both false negative and false positive detections.
Thus, optimized suites cannot be a proxy for a regular application benchmark but can provide a fast performance feedback at low cost after code changes in certain situations. 
For example, an optimized suite could be routinely run for (almost) every code change to detect most performance problems, while a more reliable application benchmark could be used as a daily backup process to detect the missed ones.

Overall, our findings open opportunities for practitioners to include new continuous benchmark steps in CI/CD pipelines and to shorten the execution times of established ones. 
Our results motivate further studies using other systems, developing further microbenchmark selection algorithms, and fine-tuning parameters to cost-efficiently improve the benchmark accuracy.
\section{Acknowledgments}
Christoph Laaber has received funding from The Research Council of Norway (RCN) under project 309642.

\bibliographystyle{IEEEtranSN}
\bibliography{bibliography}

\vfill\eject

\end{document}